\documentclass[11pt]{article}

\usepackage{graphicx}
\usepackage{amsmath}
\usepackage{mathrsfs}
\usepackage{amsfonts}
\usepackage{subfigure}
\usepackage{color}
\usepackage{multirow}

\evensidemargin \oddsidemargin

\usepackage{lineno,hyperref}
 \usepackage{graphics}

\graphicspath{{figures/}}
\usepackage{epsfig}

\newtheorem{Corollary}{\quad Corollary}

\newtheorem{proposition}{\quad Proposition}

\newtheorem{remark}{\quad Remark}

\newtheorem{theorem}{\quad Theorem}

\modulolinenumbers[1]

\begin{document}

\title{Qualitative analysis of a mathematical model for
        {\it  Xylella fastidiosa} epidemics.}

\author{
Edoardo Beretta
\thanks{CIMAB (Interuniversity Centre for Mathematics Applied to Biology, Medicine and Environment), Italy.}
\and Vincenzo Capasso
\thanks{ADAMSS (Advanced Applied  Mathematical and Statistical Sciences), Universit\`a  degli Studi di Milano, Italy.
        CIMAB (Interuniversity Centre for Mathematics Applied to Biology, Medicine and Environment), Italy.}
\and Simone Scacchi
\thanks{Dipartimento di Matematica, Universit\`a degli Studi di Milano, Italy.}
\and Matteo Brunetti
\thanks{Department of Agricultural and Environmental Sciences
    - Production, Landscape, Agroenergy,
    Universit\`a  degli Studi di Milano, Italy.}
\and Matteo Montagna
\thanks{Department of Agricultural and Environmental Sciences
    - Production, Landscape, Agroenergy,
    Universit\`a  degli Studi di Milano, Italy.}
}

\maketitle

\begin{abstract}
    In Southern Italy, since 2013, there has been an ongoing Olive Quick Decline Syndrome (OQDS) outbreak, due to the bacterium {\it Xylella fastidiosa}. In a
     couple of previous papers, the authors have proposed a mathematical approach
     for identifying possible control strategies for eliminating or at least reduce the
     economic impact of such event. The main players involved in OQDS are represented by the insect vector, {\it Philaenus spumarius}, its host plants (olive trees and weeds) and the bacterium, {\it X. fastidiosa}.   A basic mathematical model has been expressed
     in terms of a system of ordinary differential equations; a preliminary analysis
     already provided interesting results about possible control strategies within an
     integrated pest management framework, not requiring the removal of the 
     productive resource represented by the olive trees. The same conjectures have been
     later confirmed by analyzing the impact of possible spatial heterogeneities on
     controlling a {\it X. fastidiosa} epidemic.
     These encouraging facts have stimulated a more detailed and rigorous 
     mathematical analysis of the same system, as presented in this paper.
     A clear picture of the possible steady states (equilibria) and their stability
     properties has been outlined, within a variety of different parameter scenarios,
     for the original spatially homogeneous ecosystem.
     
    The results obtained here confirm,  in a mathematically rigorous way,
    what had been conjectured in the previous papers, i.e. that the removal of a
    suitable amount of weed biomass (reservoir of the juvenile stages of the insect
    vector of {\it X. fastidiosa} from olive orchards and surrounding areas is the
    most acceptable strategy to control the spread of the OQDS. In addition, as
    expected, the adoption of more resistant olive tree cultivars has been shown to
    be a good strategy, though less cost-effective, in controlling the pathogen.
\end{abstract}


\section{Introduction}

The etiological agent of the  olive quick decline
 syndrome (OQDS), a disease that have seriously affected the olive production in Apulia region (Italy) since 2013, is the plant pathogenic bacterium {\it Xylella fastidiosa} (Proteobacteria, Xanthomonadaceae).
 Once a plant is infected, bacteria multiplication within the xylem vessels  can lead to the formation of a biofilm, which can occlude the xylem vessels,
 thus inhibiting   the plant water supply. Typical symptoms are leaf scorch, dieback of twigs, branches and even of the whole plant (see e.g. \cite{carlucci2013}).

In addition to olive trees, {\it Xylella fastidiosa} can infect a large number of
 other  plants, some of which crops of relevant economic interest, such as  grapevines, almond trees, citrus plants, etc.
 (see e.g. \cite{jeger_2018}).
 
 The main vector of {\it Xylella fastidiosa} in Southern Italy has been identified
 in the so-called meadow spittlebug, i.e.,  {\it Philaenus spumarius (Hemiptera, Aphrophoridae)}, a xylem sap-
 feeding specialist (see e.g.\cite{martelli_2016}). In an olive orchard, the  juvenile form (nymphs) develops on
 weeds or ornamental plants,  in a  self-produced foam  for
 protection from predators and water loss,  the adult  moves to
 olive tree canopies  at the end of the spring/early summer, where it remains until the end of the summer, before  returning  back to weeds   for reproduction..
 
 The scope of our research is the mathematical modelling of    the dinamics of  a {\it Xylella fastidiosa} epidemics within  olive orchard agroecosystems. A sound mathematical model
 let us perform predictive analysis of the relevant components of the system, so as to suggest
 possible control strategies.
 
 In previous papers (\cite{brunetti_VK_etal_ECOMOD_2020}, \cite{anita_VK_scacchi_BMAB_2021}), motivated by the outbreak of OQDS in Southern
 Italy, models describing the epidemic have been presented.

In \cite{brunetti_VK_etal_ECOMOD_2020}, a basic model was presented, based on a system of ordinary differential
 equations (ODEs), describing a basic spatially homogeneous ecosystem includ-
 ing the main three players, i.e., the insect vector, {\it P. spumarius}, the
 olive trees, and weeds. In the same paper only a  preliminary mathematical
 analysis had been reported which anyhow anticipated rather encouraging results
 concerning satisfactory agronomic practices for the control and possible eradication of a
 {\it Xylella fastidiosa} epidemic on olive trees.
 
 We have then been motivated to carry out a more detailed and rigorous
 mathematical analysis of the same system, which has been the scope of paper
 \cite{anita_VK_scacchi_BMAB_2021}, and  of the present paper.
 
 While paper \cite{anita_VK_scacchi_BMAB_2021} had been devoted to explore the impact of possible spatial heterogeneities on controlling a {\it X. fastidiosa} epidemic, in the present paper we have finally
 succeeded in establishing  a clear picture of the possible steady states (equilibria)
 and their stability properties, within a variety of different parameter scenarios,
 though for a spatially homogeneous ecosystem. The results obtained here confirm, supported by  a mathematically rigorous analysis, what had been already conjectured in the
 previous papers, i.e. that "the removal of a significant amount of weeds (acting
 as a reservoir for juvenile insects) from olive orchards and surrounding areas  
 has resulted in the most efficient strategy to control the spread of the OQDS.
  In addition, as expected, the adoption of more resistant olive tree cultivars has
  been shown to be a good strategy, though less cost effective, in controlling the
  pathogen." \cite{brunetti_VK_etal_ECOMOD_2020}.
  
  The theoretical mathematical analysis has been supported by a set of numerical experiments, which show in a quantitative way the role of crucial parameters
  of the system for possible control, with particular attention to the choice of the olive 
  cultivar (with respect to their  resistance to {\it X. fastidiosa}  infections) and the weed elimination in the relevant orchards. It is worth mentioning that, in recent investigations presented in \cite{Schneider_2020}, the authors, by means of a
  cellular automaton simulator, have confirmed the relevance of the olive cultivar
  as a possible control strategy.

  The paper is organized as follows. In Sections \ref{materials} and \ref{mathematics} the mathematical
  model is presented. In Section \ref{equilibria} feasible equilibria have been obtained
 their stability properties have been analyzed in Section   \ref{stability}. Finally
  in Section \ref{numerics} numerical simulations are presented which confirm the analytical
  results; in the numerical simulations, the relevant parameters have been taken
  from \cite{brunetti_VK_etal_ECOMOD_2020}. The paper ends with relevant concluding remarks, in Section  \ref{Concluding remarks}.


\begin{figure}[htb]
    \centering
    \includegraphics[scale=0.25]{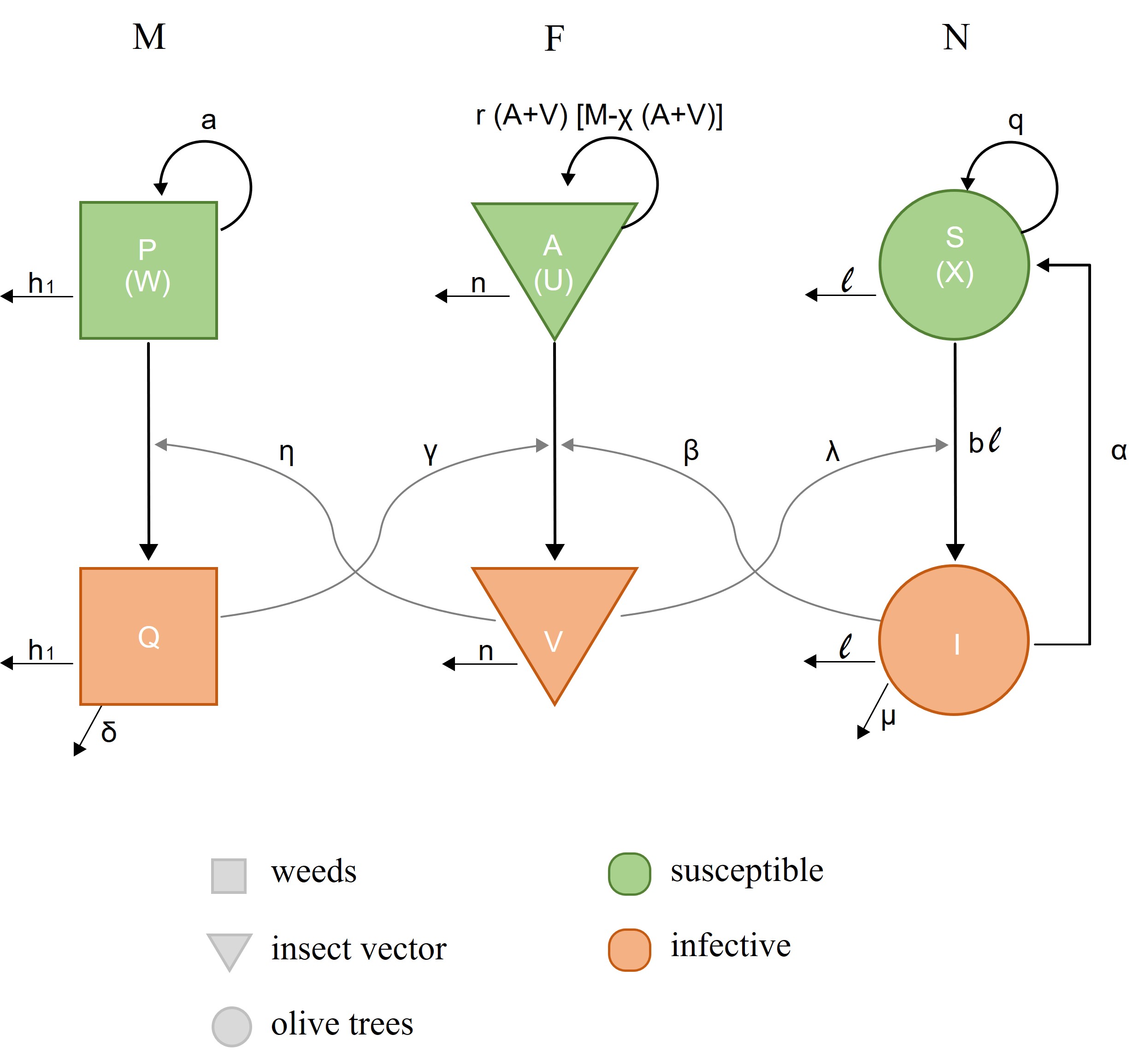}
    \caption{Diagram of interactions.}\label{Fig_X1}
\end{figure}

\section{Building blocks of the  mathematical model}   \label{materials}

As anticipated in the Introduction, we shall analyze here the same model
 proposed in  \cite{brunetti_VK_etal_ECOMOD_2020}, including what can be considered the most significant features  of the dynamics of a real epidemic system, with respect to possible control
 strategies. Accordingly only the following components have been considered.

  The individuals of the insect ({\it P. Spumarius})  population will
 be denoted by  $A$ if healthy and by $V$ if infected. The populations of susceptible and infected olive trees will be
 respectively denoted by $S$ and $I$  (see   Table \ref{tab:vars}).  As a third player of the
  modelled ecosystem we consider the so called weeds, which collectively include
  all herbaceous and shrub-like plants that may constitute a reservoir for the
  bacterial pathogen {\it X. fastidiosa}. The number of healthy weeds will be denoted
  by $P,$ while $Q$ stands for the infected ones. All the parameters in the model are
  non-negative quantities.

\subsection{The   dynamics of insects}  \label{mathematical_model_insects}

The  equations describing the evolution of
the two insect subpopulations are  the following  ones

\begin{equation} \label{adults}
\left\{
\begin{array}{l}
\displaystyle{ \frac {dA}{dt}  = r(A+V)(P+Q)  - r \chi A (A+V) - n A } \\
\quad \quad \quad \displaystyle{-\beta \frac {I}{S+I} A -\gamma \frac {Q}{P+Q} A,} \\
\displaystyle{\frac {dV}{dt} = - n  V - r \chi V (A+V)}
\displaystyle{+\beta \frac {I}{S+I} A +\gamma \frac {Q}{P+Q} A.}
\end{array}
\right.
\end{equation}

It has been taken into account the fact that bacteria are  not vertically transmitted by female insects, so that the latter generate only healthy offspring, independently
 of their status as healthy or infected (see e.g. \cite{almeida2005}, \cite{redak2004}, and references therein).
 The development of nymphs and their molting into adults, however, require weeds
 in the environment (either healthy or infected); this has been expressed by the
 dependence (here assumed to be linear) of the birth rate upon the total weed
 population. We may notice that the overall reproduction rate of insects is given by
\begin{equation}
r(A+V) [(P+Q)  -  \chi (A+V)],
\end{equation}
where a logistic term   $[(P+Q)  -  \chi (A+V)]$  has been introduced; this means that the total population of weeds  $(P+Q)$  acts as carrying capacity for the insects;
  $\chi$  has been introduced  as  a tuning parameter  with  respect  to available data.

Insects  experience a natural mortality  at a rate  $n,$  which here is assumed
 to be a constant parameter, as a technical simplification. They 
 may become infected by feeding on infected trees or plants. The insect infection
 rate is assumed to be a linear function of the relative abundances of infected
 biomasses (with respect to their respective total values) of both trees and weeds,
 via the parameters $\beta$ and $\gamma,$ respectively.

\subsection{The  dynamics of olive trees}  \label{mathematical_model_trees}

For  the olive trees  it is better  to refer to their  canopies,
so that we may consider pruning and regrowth. Their  dynamics is described by  the following two
equations:

\begin{equation} \label{trees}
\left\{
\begin{array}{l}
\displaystyle{ \frac {dS}{dt} = \left (q -  \frac {S+I}C \right ) S  - \ell S  - \lambda V  S- b \ell \frac{I}{I+S} S   + \alpha I, } \\
\quad  \\
\displaystyle{\frac {dI}{dt} =  -  \frac {S+I}{C} I  + \lambda V S
- \mu I - \ell I + b \ell \frac{I}{I+S}  S - \alpha I.}
\end{array}
\right.
\end{equation}

Healthy trees (canopy) $S$ are produced by regrowth (or additional
planting).  The  production of healthy trees has been   described
by  a logistic growth  model
\begin{equation}
\left[q - \frac{S+I}{C} \right] S
\end{equation}
where $q$   is the  natural constant  growth rate,  and  the
logistic term $ \displaystyle \frac{S+I}{C}$ takes into account a
possible carrying capacity $C.$    Correspondingly  in the second
equation, concerning  infected trees, a   logistic term
$\displaystyle \frac{S+I}{C} I$  has been  included.

For trees, in view of their long survival, natural mortality has been neglected;
 a constant decay rate $\ell,$  due to regular pruning (or possible elimination/logging)
 has instead been included. Canopies of infected trees $I$ experience a disease-
 related extra mortality  $\mu.$ A possible recovery of trees might be considered at a
 constant rate $\alpha.$
 
 Trees get infected by contact with infected adult insects, or by human ac-
 tivities such as pruning, budding and grafting, due to the use of infected tools. As far as the incidence rate due
 to infected insects is concerned the following form has been assumed, after a
 reasoning supported by \cite{dietz1982} (see also \cite{VK09}),
 
\begin{equation}
\lambda  V  S.
\end{equation}

The incidence rate due to human activities has been considered proportional
to the relative abundance of infected trees with respect to their total mass.
Given $\ell,$ the rate of contacts with tools employed for human activities, we have

\begin{equation}
b \ell \frac {I}{I+S} S.
\end{equation}

\subsection{The    dynamics of weeds}  \label{mathematical_model_weeds}

The dynamics of the weeds  mass is described by  the following two
equations:

\begin{equation} \label{plants}
\left\{
\begin{array}{l}
\displaystyle{ \frac {dP}{dt} = \left( a   -  \frac{P}{C_2} \right ) (P+Q) - \eta V P - h_P (P,Q) ,  } \\
\quad \\
\displaystyle{\frac {dQ}{dt} =  -   \frac{P+Q}{C_2}Q + \eta V P -
\delta Q - h_Q (P,Q) .}
\end{array}
\right.
\end{equation}

As  above,   logistic  growth is assumed, at a net reproduction rate
$a$ and carrying capacity $C_2;$   we assume that  all weeds
produce healthy ones. For the infection rate of weeds we have made
the same assumptions as for  the olive trees, so that the
incidence rate for weeds is

\begin{equation}
\eta   V  P.
\end{equation}

Disease-related mortality   of weeds occurs at rate $\delta,$
while  $h_P$ and $h_Q$  represent mass  reduction  due to
human-related activities. We have assumed that
they are linearly dependent on the size of the existing vegetation, i.e.:

\begin{equation}   \label{h_1}
h_P (P, Q) = h_1 P,  \quad  h_Q(P, Q) = h_1 Q.
\end{equation}

Later $h_1$ (year$-1$) will be used as a control parameter for the eventual eradi-
 cation of the epidemic in the relevant habitat;  $ h_1 = 0$  will mean that weeds are
 subject only to their natural dynamics.

 An important remark is due concerning the above model.

\begin{remark}   \label{remark_1}

 In Systems (\ref{adults}) and (\ref{trees}), the terms

 \begin{equation}   \label{remark_1_1}
\displaystyle \frac{I}{I + S} \quad  \textnormal{and}  \quad   \displaystyle \frac{Q}{P+Q} 
 \end{equation}
 may degenerate, i.e. their denominators may become zero. For a sound mathe-
matical model, in either case we have to assume that the whole fraction is taken
as zero.

\end{remark}

\begin{table} \caption{  Model variables.}
    \label{tab:vars}
    \centering
    \begin{tabular}{|c|c|}
        \hline
        Symbol & Description  \\

        \hline

        A &  Healthy insects \\

        V &  Infected insects \\

        U & Fraction of healthy insects \\

        F & Total population of insects \\

        S & Healthy olive trees  \\

        I & Infected olive trees \\

        X & Fraction of healthy olive trees \\

        N & Total canopy mass of olive trees \\

        P & Healthy weeds  \\

        Q & Infected weeds \\

        W & Fraction of healthy weeds \\

        M & Total mass of weeds\\
        \hline
    \end{tabular}
\end{table}


\begin{table} \caption{Model  parameters.}
    \label{tab:pars}
    \centering
    \begin{tabular}{|c|c|}
        \hline
        Symbol & Description  \\

        \hline

            $r$ &  Insects birth rate\\

    $\chi$ &  Insect intraspecific competition rate \\

        $n$ & Insects mortality rate \\

        $q$ &  Healthy trees (canopy) regrowth rate \\

        $C$ &  Trees carrying capacity parameter \\

        $\ell$ &  Elimination rate  of trees by pruning or logging \\

            $b$ & Infection rate  of trees  by infected tools  \\

        $\mu$ &  Infected trees mortality rate  \\

        $\alpha$ & Infected trees recovery rate  \\

        $a$ & Weeds net growth rate  \\

        $C_2$ & Weeds carrying capacity parameter  \\

        $\delta$ & Weeds mortality rate   \\

               $\beta$  &  Insects  infection rate by infected trees  \\

        $\gamma$ & Insects infection rate by infected weeds  \\

        $\lambda$ & Trees infection rate by infected insects  \\

        $\eta$ & Weeds infection rate by infected insects\\
        
         $h_1$ & Weeds elimination rate  by human intervention  \\
    \hline
    \end{tabular}
\end{table}

\section{The  model with   fractions}    \label{mathematics}
For the sake of simplicity we take all absolute populations as adimensional quantities. We will now rewrite our evolution 
equations in terms of total populations and their susceptible fractions (see also Table \ref{tab:vars}): the total number
 of insects $F = A + V,$ and the fraction of susceptibles $U=A F^{-1};$  the total
 canopy mass of olive trees $N=S+I$ and the fraction of susceptible mass
  $X=SN^{-1};$ the total weeds mass $M = P + Q$ and the fraction of susceptible
  mass $W=PM^{-1}.$    In terms of these variables our system becomes

\begin{equation} \label{adults+trees+plants}
\left\{
\begin{array}{l}
\medskip
\displaystyle{\frac {dU}{dt} = r  M (1-U) -  [\beta (1-X) + \gamma (1-W)] U , } \\
\medskip
\displaystyle{\frac {dF}{dt} = F [ r (M  - \chi F) - n ] , } \\
\medskip
\displaystyle{\frac {dX}{dt} = X [ ( q + \mu - b \ell ) (1-X) - \lambda (1-U) F ] + \alpha (1-X),  }\\
\medskip
\displaystyle{\frac {dN}{dt}  = N [ q X   - \ell  - \frac NC - \mu (1-X)], } \\
\medskip
\displaystyle{\frac {dW}{dt} =  a(1-W)+ W [\delta (1- W) - \eta (1-U) F  ] ,  } \\
\medskip
\displaystyle{\frac {dM}{dt} = M \left[ a  -  \frac{M}{C_2} -\delta  (1-W) - h_1 \right] . }
\end{array}
\right.
\end{equation}

\begin{remark}   \label{remark_2}

 As noticed above in Remark \ref{remark_1} this system may degenerate in case
 either $M$ or $F$ or $N$ becomes zero, so that it has to be complemented by the
 assumption that the corresponding fraction looses its meaning as such. For
 example, if $M = 0$ then any value of $W \in  [0, 1]$  will make $W M = 0,$   which
 is coherent with its biological meaning: if the total weed mass is zero, then the
 mass of healthy weeds is zero too.
\end{remark}

It is not difficult to show that if System (\ref{adults+trees+plants}) is subject to initial conditions
 $U(0) \in  (0, 1), F(0) > 0;  X(0) \in  (0, 1); N(0) > 0; W(0) \in (0, 1), M(0) > 0,$ then
 there exist $\bar{F} , \bar{N},  \bar{M}$ such that, for any time $t > 0, $  $U(t) \in (0, 1),  0 < F(t) <\bar{F};$   $ X(t) \in (0, 1),  0 < N(t) < \bar{N};$   $ W(t) \in (0, 1),  0 < M(t) < \bar{M}. $
 
 We shall denote  by $ \mathcal{U} := (0, 1) \times (0, \bar{F}) \times (0, 1)\times (0, \bar{N})\times (0, 1) \times (0,\bar{M} );$   $\bar{\mathcal{U}}$ shall denote the closure of $ \mathcal{U}.$
 So that we may claim that $\bar{\mathcal{U}}$ is an invariant region for System (\ref{adults+trees+plants}).

Our aim here is to analyze the qualitative behaviour of System  (\ref{adults+trees+plants}). First of  all we may look for the possible existence of equilibria, which can be obtained by solving the following system of equations, subject to Remark \ref{remark_2},
\begin{eqnarray}
& & r  M (1-U) -  [\beta (1-X) + \gamma (1-W)] U = 0, \label{eq_1}  \\
& & F [ r (M  - \chi F) - n ] =0 ,  \label{eq_2}\\
& & X [ ( q + \mu - b \ell ) (1-X) - \lambda (1-U) F ] + \alpha (1-X) = 0,  \label{eq_3}\\
& & N \left[ q X   - \ell  - \frac NC - \mu (1-X) \right] = 0,  \label{eq_4} \\
& &  a(1-W)+ W [\delta (1- W) - \eta (1-U) F  ] = 0,  \label{eq_5} \\
&  & M \left[ a  -  \frac{M}{C_2} -\delta  (1-W) - h_1 \right] = 0 . \label{eq_6}
\end{eqnarray}

\section{Equilibria} \label{equilibria}

Consider first the case $a > 0$ and $h_1 = 0.$

 We will carry out the analysis of the possible equilibria of System (\ref{eq_1})-(\ref{eq_6})
 in terms of the value of $\lambda,$ the infection rate of olive trees by infective insects,
 which expresses the resistance to infection by a specific cultivar.

 \subsection{ The disease free equilibrium}

  For $\lambda = 0,$   Equation (\ref{eq_3}) can be rewritten as
  
  \begin{equation}  \label{eq_3_1}
  X[(q + \mu - b \ell)(1 - X)] + \alpha (1 - X) = 0,  
 \end{equation}
 which admits the solution $X_1 = 1.$

  From here Equation (\ref{eq_4}) becomes
  
  \begin{equation}    \label{eq_4_1}
\displaystyle  N \left [q - \ell - \frac{N}{C}    \right]     =0.
 \end{equation}

  This admits the solution   $ N_1 = C(q -  \ell),$  which is biologically intuitive.
  
  If in addition $\eta = 0,$ Equation (\ref{eq_5}) becomes

  \begin{equation}   \label{eq_5_1}
 a(1- W ) + W [\delta(1 -  W )] = 0,  
 \end{equation} 
  which admits the solution $ W_1 = 1.$
  
  As a consequence Equation (\ref{eq_6}) becomes
  
  \begin{equation}    \label{eq_6_1}
\displaystyle     M \left[a- \frac{M}{C_2}     \right] =0,  
\end{equation}  
which admits the solution  $ M_1 = C_2 a.$

If $\displaystyle  C_2 a > \frac{n}{r},$    then Equation (\ref{eq_2}) admits the solution
 \begin{equation}    \label{eq_2_1}
\displaystyle     F_1 =  \frac{1}{\chi}\ \left[M_1 - \frac{n}{r}     \right].  
\end{equation}

 Moreover Equation (\ref{eq_1}) becomes
 
  \begin{equation}    \label{eq_1_1}
rM(1 -  U) = 0,  
 \end{equation}  
 which admits the solution $U_1 = 1.$

  To conclude, in absence of transmission, i.e. for $\lambda  = \eta =0,$ if we make the
trivial assumptions that $q > \ell,$ and $ \displaystyle C_2 a >\frac{n}{r},$
 which are satisfied as from Table  \ref{tab:param},  it is not difficult to check that the following one is a nontrivial equilibrium of   the ODE system (\ref{adults+trees+plants})
 
  \begin{equation}    \label{disease_free_equilibrium}
\displaystyle  E_1 = (U_1,  F_1, X_1, N_1 W_1 M_1) := (1, \frac{1}{\chi}\ \left[M_1 - \frac{n}{r}     \right] 1; C(q - \ell), 1; C_2 a).  
 \end{equation}

 It is clear that we may obtain the same equilibrium by imposing a priori
 $U = 1;$  this situation  is anyhow less interesting from the point of view of the stability
 analysis.

 \begin{table} \caption{ Values of the model parameters. For parameters given in a range,
 		the value before the brackets has been used for simulations, unless  otherwise specified.}
 	\label{tab:param}
 	
 	\centering
 	
 	\begin{tabular}{|c|c|c|}
 		\hline
 		Symbol  & Values & References \\
 		
 		\hline
 		
 		$r$  & 200 [37,400] year$^{-1}$ & \cite{silva2015,Yurtsever_2000}  \\
 		
 		$\chi$  &  0.001, 0.01 & \cite{brunetti_VK_etal_ECOMOD_2020} \\
 		
 		$n$ & 0.98 [0.95, 0.99] year$^{-1}$ & \cite{brunetti_VK_etal_ECOMOD_2020} \\
 		
 		$q$  & 0.5 [0.2, 0.7] year$^{-1}$ & \cite{Villalobos_2006}  \\
 		
 		$C$  &  100  & \cite{brunetti_VK_etal_ECOMOD_2020} \\
 		
 		$\ell$  & 0.01 year$^{-1}$ & \cite{brunetti_VK_etal_ECOMOD_2020} \\
 		
 		$b$  & $ 0.05$ & \cite{brunetti_VK_etal_ECOMOD_2020} \\
 		
 		$\mu$  & 0.9 [0.8, 1] year$^{-1}$ & \cite{Saponari_2017, Saponari_2018}  \\
 		
 		$\alpha$  & 0.1, 0.5 year$^{-1}$ & \cite{brunetti_VK_etal_ECOMOD_2020} \\
 		
 		$a$  & 0.3 [0.1, 1.] year$^{-1}$ & \cite{brunetti_VK_etal_ECOMOD_2020} \\
 		
 		$C_2$  & $10 \times C$ & \cite{brunetti_VK_etal_ECOMOD_2020} \\
 		
 		$\delta$  &  0.2 [0, 0.5] year$^{-1}$ & \cite{Saponari_2018}  \\
 		
 		$h_1$  & [0, 0.8] year$^{-1}$ & \cite{brunetti_VK_etal_ECOMOD_2020} \\
 		
 		$\beta$   & 0.75 year$^{-1}$ & \cite{cornara2017} \\
 		
 		$\gamma$  & 0.1 [0.1, 0.5] year$^{-1}$ & \cite{cornara2017} \\
 		
 		$\lambda$  & [0.2, 0.8] year$^{-1}$ & \cite{Boscia_2017, Fierro_2019} \\
 		
 		$\eta$ & 0.1 [0.1, 0.6] year$^{-1}$ & \cite{cornara2017} \\
 		\hline
 	\end{tabular}
 \end{table}

\subsection{Other equilibria}  \label{other_equilibria}

 Let us now consider the case $\lambda > 0,$ and look for nontrivial equilibria, by im-
 posing that the ecosystem is exposed to a nontrivial infective insect population,
 which is given by $(1 - U)F > 0.$
 Consider first the equation for  $ X,$ the fraction of 
 healthy olive tree biomass
 
   \begin{equation}  \label{eq_3_2}
 X[(q + \mu - b \ell)(1 - X) - \lambda (1 - U)F] + \alpha (1 - X) = 0,  
 \end{equation}
 
 This can be rewritten as a second order algebraic equation

  \begin{equation}  \label{eq_3_3} 
A X^2 + B X + C = 0,
 \end{equation}
with

 \begin{equation}  \label{eq_3_4} 
A := q + \mu - b \ell,
 \end{equation}

  \begin{equation}  \label{eq_3_5} 
 C := -\alpha,
 \end{equation}

    \begin{equation}  \label{eq_3_6} 
  B := \lambda (1 - U)F - (A + C).
  \end{equation}
  
 The solutions of (\ref{eq_3_3}) are given by

 \begin{equation}  \label{eq_3_7} 
\displaystyle X^*_{\pm} =   \frac{-B \pm \sqrt{B^2 - 4 AC}}{2A}.
\end{equation}

Since, by the parameter values taken from Table \ref{tab:param}, it is $A > 0,$ we may claim
  that $AC < 0,$ so that  $\sqrt{B^2 - 4AC} > B,$ and finally the existence of a nontrivial
  solution  $ X^* > 0.$

 We may further notice that $ X^* < 1$ iff  $\sqrt{B^2 - 4AC} < 2A+ B,$ i.e. iff
 $ B + A + C =  \lambda (1 - U)F > 0.$
 
 We have thus proven the following result.

 \begin{proposition}   \label{proposition_1}
 For $ \lambda > 0$ and any value of the infective insect population
 $ (1 - U)F > 0$, there exists a unique nontrivial equilibrium $ X^* \in (0, 1)$ for the
 fraction of healthy olive trees.
\end{proposition}

We may now turn to the analysis of the equilibrium equation for the total
 tree biomass, under the assumptions of the above proposition.
 
 If we look for nontrivial solutions of the equilibrium equation for $N,$
 
 \begin{equation}  \label{eq_4_11} 
\displaystyle  N \left[q X  - \ell - \frac{N}{C} - \mu (1-X)  \right] =0,
\end{equation} 
 it reduces to
 
 \begin{equation}  \label{eq_4_12} 
\displaystyle  q X  - \ell - \frac{N}{C} - \mu (1-X)  =0,
\end{equation} 
 from which we obtain the equilibrium

  \begin{equation}  \label{eq_4_13} 
 \displaystyle  N^* = C (q+\mu) \left[   X^* - \frac{\mu + \ell}{\mu + q} \right].
 \end{equation}

 This shows that (\ref{eq_4_11}) admits a solution $N^* > 0$ iff
 
 \begin{equation}  \label{eq_4_14} 
\displaystyle   X^* > \frac{\mu + \ell}{\mu + q}.
\end{equation}

 Usually   $\ell < q,$ so that we may claim that the following statement holds true.

 \begin{Corollary}   \label{corollary_1}
 Under the assumptions of Proposition  \ref{proposition_1}, a unique nontrivial equilibrium $N^* > 0$ exists for the total tree biomass iff
\begin{equation}  \label{eq_4_15} 
\displaystyle   X^* \in \left( \frac{\mu + \ell}{\mu + q} , 1\right)  \neq \emptyset.
\end{equation} 

Otherwise  (\ref{eq_4_11}) admits only the trivial solution   $N^* = 0.$
\end{Corollary}

   \begin{remark}    \label{remark_3}
 Notice that Condition (\ref{eq_4_15} ) may   apply  only  if the pruning rate $\ell$ is  smaller than the natural growth rate $q$ of the biomass.
 If  this condition does not hold then  the olive trees will eventually disappear, independently of other conditions.
   \end{remark}

 To conclude the analysis of a possible nontrivial equilibrium, let now consider
  the equilibrium equation for the fraction of healthy weeds
 
 \begin{equation}  \label{eq_5_11} 
  a(1 - W ) + W [\delta(1 - W ) - \eta(1 - U)F ] = 0.
 \end{equation}

  Under the assumption that $a > 0,$ we may introduce the quantities
  
  \begin{equation}  \label{eq_5_12} 
 \displaystyle  D:= \frac{\delta}{a},
 \end{equation}

 \begin{equation}  \label{eq_5_13} 
  \displaystyle  G:=  1- \frac{\delta}{a} + \frac{\eta}{a} (1 - U)F,
 \end{equation}

 \begin{equation}  \label{eq_5_14} 
 L=-1.
 \end{equation} 
   so that Equation (\ref{eq_5_11}) can be rewritten in the form
   
   \begin{equation}  \label{eq_5_15} 
  D W^2 + G W + L = 0.
  \end{equation}

 It admits two solutions given by
 
  \begin{equation}  \label{eq_5_16} 
 \displaystyle W^*_{\pm} =   \frac{-G \pm \sqrt{G^2 + 4 D}}{2D}.
  \end{equation}  
 
 It is not difficult to see that $D \in (0, 1)$ (which is usually the case) implies
 $ G > 0,$ so that Equation (\ref{eq_5_11}) admits a unique nontrivial solution $W^*_{+} > 0.$  
  We   may then claim that the following statement holds true.

 \begin{proposition}  \label{proposition_2} 
 Under the assumptions of Proposition \ref{proposition_1}, if further $\displaystyle \frac{\delta}{a} \in (0, 1),$
  there exists a unique nontrivial equilibrium solution for the fraction of healthy
  weeds, given by
 \begin{equation}  \label{eq_5_17} 
\displaystyle W^* =   \frac{-G + \sqrt{G^2 + 4 D}}{2D}.
\end{equation}  
\end{proposition}

 We may notice that $ \displaystyle 1+ \frac{\eta}{a} (1 - U)F > 1$ is equivalent to $D + G > 1,$ and   this is equivalent to state that $W^* < 1.$  Altogether we may then state that,
  under the assumptions of Proposition \ref{proposition_2} we have $W^* \in  (0, 1).$

  For the total weed mass the equilibrium equation is
  
   \begin{equation}  \label{eq_6_11} 
  \displaystyle  M \left[a -  \frac{M}{C_2} - \delta  (1- W)  \right] =0.
  \end{equation} 
  
 As from the analysis of the equilibrium $E_1,$  we obtain that the nontrivial
  solution of Equation (\ref{eq_6_11}) is given by
  
   \begin{equation}  \label{eq_6_12} 
  M = M^* := C_2[a -  \delta (1 - W^*)] \ge C_2(a - \delta).
  \end{equation} 
  
 Given the values of the parameters, as from Table \ref{tab:param}, we may state that
 
   \begin{equation}  \label{eq_6_13} 
 \displaystyle  M^* >   \frac{n}{r}.
 \end{equation} 
 
 Under this condition, Equation (\ref{eq_2}) admits the nontrivial solution
 
   \begin{equation}  \label{eq_6_14} 
 \displaystyle  F= F^* :=   \frac{1}{\chi} \left[ M^*  - \frac{n}{r} \right].
 \end{equation} 

 Finally we can identify the nontrivial solution of Equation (\ref{eq_1}) in the fol-
  lowing form

   \begin{equation}  \label{eq_1_11} 
  \displaystyle  U^* = M^* \left[ M^* +  \frac{\beta}{r} (1- X^*)  + \frac{\gamma}{r} (1- W^*) \right]^{-1},
  \end{equation} 
  
 All the above leads to a nontrivial equilibrium

  \begin{equation}  \label{equilibrium_E_2} 
 E_2 = (U^*, F^*,  X^*, N^*, W^*,  M^*) \in  \mathcal{U}.
 \end{equation} 
 
 We may recollect all the above analysis in the following statement.

\begin{proposition}
\label{proposition_3} Under the assumption that $ \displaystyle a > \delta > 0,  C_2[a - \delta] > \frac{n}{r} , $
 and $ \ell < q,$ a nontrivial equilibrium $E_2 = (U_2, F_2, X_2, N_2, W_2, M_2)$ may exist for System
  (\ref{adults+trees+plants}) in the open domain $\mathcal{U},$ provided Condition (\ref{eq_4_15} ) is satisfied. Otherwise this
  equilibrium degenerates into $ E_3 = (U_3, F_3, X_3, N_3, W_3, M_3) $ with

   \begin{equation}  \label{equilibrium_E_3} 
\displaystyle  U_3 \in (0,1), F_3 =  \frac{1}{\chi} \left(  M_3 -  \frac{n}{r}   \right),  X_3 \in \left(0, \frac{\mu + \ell}{\mu+ q}\right), N_3 =0, W_3 \in (0, 1),  M_3 > \frac{n}{r}.
  \end{equation} 
\end{proposition}

   \subsection{The  role of $\lambda$}
   
   From an agronomic point of view, it is interesting to discuss about
 the dependence of the possible equilibrium $X^*,  N^*$ upon  $\lambda$ (in a similar way one
  might think about the role of $\eta$ too, but this is practically  impossible to control by usual
 agronomic practices, since $\eta$ represents the resistance of weeds to contagion). Up
 to now we have noticed that for $\lambda = 0$ - absence of contagion to olive trees -
  $X^* = 1$  and $N^* = C(q - \ell),$  as in the equilibrium $E_1.$
  
  The role of $\lambda$ in the case   $\lambda > 0$ is in general more difficult to analyze. But
  if we restrict ourselves to the case $\alpha  = b = 0, $ (\ref{eq_3_3} ) simply becomes
  
    \begin{equation}  \label{49} 
  A X^2 + B X = 0 ,
  \end{equation}
  with   $A=q+\mu,$  $B=\lambda (1-U)F - A,$ and $C=0.$

 This equation admits, in addition to the trivial solution, the nontrivial solution
   \begin{equation}  \label{50} 
 \displaystyle  X^* =  \frac{-B}{A}
 \end{equation} 
  i.e.
  
 \begin{equation}  \label{51} 
\displaystyle  X^* = 1-  \frac{\lambda}{q+ \mu} (1-U) F
\end{equation} 
  which finally becomes
 \begin{equation}  \label{52} 
 \displaystyle  X^* = 1-  \frac{\lambda}{q+ \mu} (1-U^*) F^*,
 \end{equation} 
 if we take the corresponding equilibrium values for $U$ and $F.$
 
  This expression shows that the value of $X^*$ decreases as $\lambda$ increases, as it might be conjectured. 
  
  Let us   investigate the impact of this result on the total olive tree biomass.
  
  For the equilibrium $E_2,$ the equilibrium value $N_2$ satisfies Equation (\ref{eq_4_13})  that
 we report here
 
  \begin{equation}  \label{53} 
 \displaystyle  N_2 = C (q+\mu) \left[   X_2 - \frac{\mu + \ell}{\mu + q} \right].
 \end{equation}

  If we impose that $N_2 > 0,$  by Equation (\ref{52} ) we have to impose
 
  \begin{equation}  \label{54} 
 \displaystyle  N_2 = C (q+\mu) \left[   X_2 - \frac{\mu + \ell}{\mu + q} \right] = C \left[q - \ell  - \lambda (1- U_2)F_2\right] > 0 
 \end{equation} 
 which requires

  \begin{equation}  \label{55} 
 \displaystyle   \lambda < \frac{q- \ell}{(1- U_2)F_2 }. 
 \end{equation} 
 
 On the other hand the equilibrium $E_3$ corresponds to the case in which
$\displaystyle X_3 < \frac{\mu+\ell}{\mu + q}$   i.e., by  Equation (\ref{52}),

 \begin{equation}  \label{56} 
\displaystyle   \lambda > \frac{q- \ell}{(1- U_2)F_2 }. 
\end{equation}

  These two inequalities (\ref{55}) and (\ref{56}) shed some light on the role of $\lambda$ in case
  of an epidemic, which means a nontrivial value of the infective insect population
 $ (1 - U^*)F^*,$  at equilibrium: for a sufficiently small $\lambda$ we may have the equilibrium $E_2,$ i.e. the coexistence of a nontrivial olive tree biomass. 
 This is not possible for sufficiently large values of  $\lambda,$ in which case only the equilibrium $E_3$
  is feasible, which means extinction of the olive tree biomass.
  
 Actually a rigorous reasoning should take into account that the quantity
  $ (1 - U^*)F^*$   may depend itself upon $\lambda.$
  
 Anyhow, the above discussion has been confirmed by the numerical simulations  (see Figures  \ref{fig_E_2}, \ref{fig_E_4} ): the choice
 of more resistant cultivar may lead to coexistence.
 
 This is a practice already implemented in Southern Italy. In
 an optimal control problem, it has to be compared with quality and yield of
  more resistant cultivar, with respect to less resistant ones (see Section    \ref{Concluding remarks} for the
 concluding remarks).

 In the following we shall investigate a different practice, which does not
 impose change of the  olive tree cultivar, by acting instead on the agronomic practice of eliminating (or at least significantly reduce )  weeds  in the relevant orchards.

 \subsection{The case  $h_1>a$}
 
 Let us then analyze the case $a > 0,$ with $h_1 > a,$ according to which the
  weed mass cannot increase, and eventually dies out.
  
  In fact, under this assumption, the only feasible solution of Equation (\ref{eq_6}) is
  the trivial one  $M_4 = 0.$  As a consequence, from Equation (\ref{eq_2}), the only possible
  equilibrium for  $ F$ is the trivial one $F_4 = 0.$ By taking into account Remark   \ref{remark_2}, this implies that any values $W_4 \in [0, 1] $ and $ U_4 \in [0, 1]$ are admissible, hence  irrelevant for further analysis. Moreover, from Equation (\ref{eq_3}),$ X_4 = 1$ so that,  from Equation (\ref{eq_4}), we obtain $N_4 = C(q -  \ell).$
  
 We may then conclude with the following proposition.

\begin{proposition}  \label{proposition_4}
	Under the assumption that $a > 0,$    $\ell < q$  and $ h_1 > a,$  System
 (\ref{adults+trees+plants}) admits the following equilibrium
 
  \begin{equation}  \label{57} 
E_4 = (U,  0,  1,  C(q - \ell), W, 0), 
 \end{equation} 
for irrelevant values of $U$ and $W,$ which stay anyway bounded in  $ [0, 1].$ 
\end{proposition}

  In synthesis, the equilibrium $E_1$ corresponds to a disease free ecosystem;
  equilibrium $E_2$ corresponds to coexistence of a nontrivial olive tree biomass
  and infective insects, which we may conjecture is possible only for a sufficiently
  small value of $\lambda,$  i.e. for more resistant olive tree   cultivars; the equilibrium $E_2$ may
  degenerate into $E_3$ for a sufficiently large value of $\lambda,$ i.e. for less resistant olive
  cultivar. Finally the equilibrium $E_4$ corresponds to the eradication of the insect
  population, induced by the eventual eradication of the weed biomass.

\section{Stability}      \label{stability}   

\subsection{Stability of the equilibria $E_2$ and $E_3$}  \label{stability_E2_E3}

 We shall consider first the case of the existence of a nontrivial equilibrium
$ E_2 = (U_2, F_2, X_2,  N_2, W_2,  M_2)$ for System (\ref{adults+trees+plants}) in the open domain
 $ \mathcal{U}. $
 
  We remind here that $X_2$ denotes the positive solution of the equation
  
   \begin{equation}  \label{58} 
A X^2 + B X + C = 0, 
 \end{equation}  
 with
 
  \begin{equation}  \label{59} 
 A := q + \mu - b \ell, 
 \end{equation} 
 
  \begin{equation}  \label{60} 
 C := -\alpha, 
 \end{equation} 
 and
 
   \begin{equation}  \label{61} 
 B := \lambda (1 - U_2)F_2 - (A + C).
 \end{equation}

We shall denote the negative solution of (\ref{58}) by $X^-_2.$

 On the other hand we denote by $W_2$ the positive solution of the
equation

  \begin{equation}  \label{62} 
D W^2 + G W + L = 0, 
\end{equation}  
where

  \begin{equation}  \label{63} 
\displaystyle  D:= \frac{\delta}{a},
\end{equation}

\begin{equation}  \label{64} 
\displaystyle  G:=  1- \frac{\delta}{a} + \frac{\eta}{a} (1 - U_2)F_2,
\end{equation}

\begin{equation}  \label{65} 
L=-1.
\end{equation}

 We shall denote the negative solution of (\ref{62}) by $W_2^-.$

Let us then introduce the functions

\begin{equation}  \label{66} 
\Lambda (X):=  A(X - X_2^- ),  \quad  \textnormal{for} \quad X  \in [0, 1],
\end{equation} 
and

\begin{equation}  \label{67} 
\Gamma (X):=  D(W - W_2^-),  \quad  \textnormal{for} \quad W  \in [0, 1].
\end{equation} 

 It is clear that $\Lambda (X) > 0,$ for any  $X \in  [0, 1],$  and $ \Gamma (W) > 0,$ for any $W \in [0, 1].$
 
 From now on we shall denote ${\bf Z}(t) := (U(t), F(t), X(t), N(t), W(t), M(t))^T .$

 By centering System (\ref{adults+trees+plants}) with respect to the coordinates of $E_2,$ we obtain

\begin{equation}  \label{68} 
\displaystyle \frac{d}{d\,t} {\bf Z}(t) = {\bf f}({\bf Z}(t)),
\end{equation} 
for

\begin{equation}
\label{eq_69}
{\bf f}({\bf Z})=
\left(
\begin{array}{c}
\displaystyle -r\frac{M_2}{U_2}(U-U_2) \vspace{0.2cm}\\
\displaystyle -r\chi F(F-F_2) \vspace{0.2cm}\\
\displaystyle -\Lambda(X)(X-X_2) \vspace{0.2cm}\\
\displaystyle -\frac{N}{C}(N-N_2) \vspace{0.2cm}\\
\displaystyle -a\Gamma(W)(W-W_2) \vspace{0.2cm}\\
\displaystyle a\left\{M\left[-\frac{M-M_2}{a C_2} + \frac{\delta}{a}(W-W_2)\right]\right\}
\end{array}
\right), \quad {\bf Z}\in\mathcal{U}.
\end{equation}


Consider the function

\begin{equation}  \label{70} 
g (y) = y - 1 - \ln y,  \quad  y \in (0, +\infty).
\end{equation}

 It is clear that $g \in  C^1((0, +\infty))$  and it is such that  $g(y) \ge 0,$ for all $y \in
(0, +\infty),$  and $g(y) = 0$  iff $y = 1.$

Moreover $ \displaystyle g'(y) = 1 - \frac{1}{y},$   so that $g'(y) < 0$  for  $ y \in  (0, 1),$    $g'(y) > 0$ for $y \in (1, +\infty), $  and  $g'(y) = 0$  iff $y = 1.$

 In order to analyze the stability of the equilibrium $E_2,$ we take as Lyapunov function


\begin{equation}
\label{eq_71}
\begin{array}{lll}
V({\bf Z})	& :=	& \displaystyle \frac{1}{2r}[\alpha_U (U-U_2)^2 + \beta_F (F-F_2)^2] \vspace{0.2cm}\\
&	& \displaystyle +\frac{1}{2} [\alpha_X (X-X_2)^2 + \beta_N (N-N_2)^2] \vspace{0.2cm}\\
&	& \displaystyle +\frac{1}{a} \left[\alpha_W g \left(\frac{W}{W_2}\right) + \beta_M g \left(\frac{M}{M_2}\right)\right],
\end{array}
\end{equation}

where $   \alpha_U , \alpha_X, \alpha_W$ and $ \beta_F , \beta_N , \beta_M $  are positive constants to be suitably chosen.

 As a consequence of the definition, and the cited properties of the function $ g, $ it is  $V  \in C^1(\mathcal{U})$  and
 
 \begin{equation}  \label{72} 
V({\bf Z}) \ge 0 \quad \textnormal{for all} \quad {\bf Z} \in  \mathcal{U},
 \end{equation}

  \begin{equation}  \label{73} 
  V({\bf Z}) = 0 \quad \textnormal{iff} \quad {\bf Z} =  E_2.
  \end{equation} 
  
 Moreover the derivative of $V$ along the trajectories of System (\ref{68} ) is given
by


{\small
\begin{equation}
\label{eq_74}
\begin{array}{lll}
& & DV({\bf Z}(t))	 :=	 \displaystyle \mbox{grad} V({\bf Z}(t))\cdot\frac{d}{dt}{\bf Z}(t) = \mbox{grad}V({\bf Z}(t)) \cdot {\bf f}({\bf Z}(t)) = \vspace{0.2cm}\\
&	& \displaystyle -\left[\alpha_U\frac{M_2}{U_2}(U(t)-U_2)^2 + \beta_F \chi F(t) (F(t)-F_2)^2\right] + \vspace{0.2cm}\\
&	& \displaystyle -\left[\alpha_X\Lambda(X(t))(X(t)-X_2)^2+\beta_N\frac{N(t)}{aC}(N(t)-N_2)^2\right]+ \vspace{0.2cm}\\
&	& \displaystyle +\left\{-\alpha_W\left(\frac{\Gamma(W(t))}{W(t)W_2}\right)(W(t)-W_2)^2+ \right. \vspace{0.2cm}\\
&	& \displaystyle \left. -\beta_M\frac{1}{a C_2 M_2}(M(t)-M_2)^2+\beta_M\left(\frac{\delta}{a M_2}\right)(M(t)-M_2)(W(t)-W_2)\right\}.
\end{array}
\end{equation}
}
The  term  within $\{...\}$  in the  above expression can be written as the  following quadratic form 

{\small
\begin{equation}
\label{eq_75}
(W(t)-W_2,M(t)-M_2)
\left(
\begin{array}{cc}
\displaystyle -\alpha_W \frac{\Gamma(W(t))}{W(t)W_2}	& \displaystyle \frac{1}{2}\frac{\beta_M}{M_2}\frac{\delta}{a} \vspace{0.2cm}\\
\displaystyle \frac{1}{2}\frac{\beta_M}{M_2}\frac{\delta}{a}	& \displaystyle -\frac{\beta_M}{M_2 a C_2}
\end{array}
\right)
\left(
\begin{array}{c}
W(t)-W_2 \vspace{0.2cm}\\
M(t)-M_2
\end{array}
\right)
\end{equation}
}

associated  with  the real symmetric matrix  
\begin{equation}
\label{eq_76}
Q=
\left(
\begin{array}{cc}
\displaystyle -\alpha_W \frac{\Gamma(W)}{W W_2}    & \displaystyle \frac{1}{2}\frac{\beta_M}{M_2}\frac{\delta}{a} \vspace{0.2cm}\\
\displaystyle \frac{1}{2}\frac{\beta_M}{M_2}\frac{\delta}{a}    & \displaystyle -\frac{\beta_M}{M^* a C_2}
\end{array}
\right).
\end{equation}

Let us examine   the structure of the matrix $Q.$

The    trace  of $Q$  is given by

\begin{equation}
\label{eq_77}
\textnormal{tr}\, Q =  -\alpha_W \frac{\Gamma(W)}{W W_2}     -\frac{\beta_M}{M_2 a C_2}.
\end{equation}

Since  both  $\alpha_W$  and  $\beta_M$ are  positive   constants, it is clear  that

\begin{equation}
\label{eq_78}
\textnormal{tr}\, Q <0.
\end{equation}

The   determinant of  $Q$  is given by 

\begin{equation}
\label{eq_79}
 \displaystyle  \textnormal{det}\, Q =  \alpha_W \frac{\Gamma(W)}{W W_2} \frac{\beta_M}{M_2 a C_2}  
 -  \frac{1}{4} \left(  \frac{\beta_M }{M_2}\frac{\delta}{a} \right)^2.
\end{equation}

We  may choose the positive constants $\alpha_W$  and  $\beta_M$  in such a way that

\begin{equation}
\label{eq_80}
\displaystyle  \frac{\beta_M }{M_2}\frac{\delta}{a}  =1  \quad  \textnormal{and} \quad 
\alpha_W \frac{1}{ W_2 a C_2} >1, 
\end{equation}

so that

\begin{equation}  \label{81} 
\det Q > 0.
\end{equation} 

 Conditions (\ref {eq_78}) and (\ref {81}) make $Q$ a stability matrix, which implies that the
 quadratic form (\ref {eq_75}) is negative definite.
 As a consequence
 
  \begin{equation}  \label{82} 
D V({\bf Z}) \le 0 \quad \textnormal{for all} \quad {\bf Z} \in  \mathcal{U},
 \end{equation} 
 
and 
 
 \begin{equation}  \label{83} 
 D V({\bf Z}) = 0 \quad \textnormal{iff} \quad {\bf Z} =  E_2.
 \end{equation}

We may then claim that the following theorem holds true.

\begin{theorem}   \label{Theorem_1}
 Under the assumptions $a > 0,$   $\displaystyle  C_2(a - \delta) > \frac{n}{r}, $
 and $   \lambda $  sufficiently small so that $\displaystyle  X_2   \in   \left(   \frac{\mu + \ell}{\mu + q}, 1 \right) \neq \emptyset, $   the equilibrium $E_2$ is globally asymptotically stable in $\mathcal{U}.$
\end{theorem}

What can we say in case the condition  $\displaystyle   X_2  >      \frac{\mu + \ell}{\mu + q}$
  does not hold? In this case the equilibrium  $E_2$   degenerates into $E_3$  in which the total olive tree mass $N$  admits only the trivial equilibrium  $N_3 = 0.$
  
Under   these  circumstances  it is more convenient to split the stability analysis of $E_3,$  by
considering on one side the stability with respect to the variables $U, F, W, M,$  and on the other side the stability with respect to the variable $ N.$ As from
 Remark \ref{remark_2}, the stability analysis of the system with respect to the variable $X$ is
 irrelevant.

For the variables $U, F, W, M,$ we may take as Lyapunov function


\begin{equation}
\label{eq_71}
\begin{array}{lll}
V({\bf Z})	& :=	& \displaystyle \frac{1}{2r}[\alpha_U (U-U_3)^2 + \beta_F (F-F_3)^2] \vspace{0.2cm}\\
&	& \displaystyle +\frac{1}{a} \left[\alpha_W g \left(\frac{W}{W_3}\right) + \beta_M g \left(\frac{M}{M_3}\right)\right],
\end{array}
\end{equation}

where $   \alpha_U , \alpha_W$ and $ \beta_F , \beta_M $  are positive constants, and proceed as above.

For the variable $N$  we may realize that its evolution equation can be written
 as
 
\begin{equation}  \label{85} 
\displaystyle \frac{d}{d\,t} N (t) =  -  \frac{N(t)}{C} (N(t)-\widehat{ N}),
\end{equation}  
for

\begin{equation}  \label{86} 
\displaystyle \widehat{ N} =  C(q + \mu) \left[   X_3 - \frac{\mu + \ell}{\mu + q}\right].
\end{equation} 

It is clear that, under the condition   $\displaystyle    X_3 <  \frac{\mu + \ell}{\mu + q},$
 the quantity  $\widehat{ N} < 0,$  so that
 
 \begin{equation}  \label{87} 
 \displaystyle \frac{d}{d\,t} N (t) =  -  \frac{N(t)}{C} (N(t)-\widehat{ N}) \le 0, \quad \textnormal{for} \quad  N(t)\ge 0
 \end{equation}  
and

 \begin{equation}  \label{88} 
\displaystyle \frac{d}{d\,t} N (t) =  -  \frac{N(t)}{C} (N(t)-\widehat{ N}) = 0, \quad \textnormal{for} \quad  N(t) = 0,
\end{equation}  
 which provides the stability of $N_3 = 0.$

 We may then state the following
 
 \begin{theorem}   \label{Theorem_2}
 	Under the assumptions $a > 0,$   $h_1 = 0,$  $\displaystyle  C_2(a - \delta) > \frac{n}{r}, $
 	and $   \lambda $  sufficiently large  so that $\displaystyle  X_2  <  \frac{\mu + \ell}{\mu + q},$   the equilibrium $E_3$ is globally asymptotically stable in $\mathcal{U} \cup \{ N=0 \}.$
 \end{theorem}

 \subsection{Stability of the disease free equilibrium $E_1$}   \label{stability_E_1}
 
 In a sense, the disease free equilibrium 
 
  $$ \displaystyle  E_1 = \left(   1, \frac{1}{\chi} \left[C_2a  - \frac{n}{r}      \right], 1, C(q- \ell), 1, C_2 a\right)$$
    is a particular case of the nontrivial equilibrium  $E_2,$ but for the fact
  that we know $ U_1 = 1, $ so that  $ (1 - U_1)F_1 = 0.$  This implies that the quantities $X_1^-$  and $W_1^-,$  
 respectively defined in (\ref{eq_3_7}) and (\ref{eq_5_16}), are given by   $ \displaystyle - \frac{\alpha}{A}$   and  $ \displaystyle - \frac{a}{\delta},$  respectively. As a consequence the quantities  $  \Lambda(X)$ and $ \Gamma(W ),$  defined respectively as  in
  (\ref{66}) and (\ref{67}), in this case are given by

  \begin{equation}  \label{89} 
 \displaystyle \Lambda (X)=  A(X  + \frac{\alpha}{A} )
  \end{equation} 
  and
  
  \begin{equation}  \label{90} 
\displaystyle  \Gamma (X)=    \frac{\delta}{a}(W +  \frac{a}{\delta} ). 
  \end{equation}

 Apart from these specifications, the stability analysis of $E_1$ can be carried
 out along the same lines as for $E_2,$ leading us to state the following
 
  \begin{theorem}   \label{Theorem_3}
 	Under the assumptions $a > 0,$   $\displaystyle  C_2 a  > \frac{n}{r}, $
 	and $   \lambda =  \eta =0,    $    the equilibrium $E_1$ is globally asymptotically stable in $\mathcal{U} \cup \{ U=1 \}  \cup \{ X=1 \}  \cup \{ W=1 \}.$
 \end{theorem}


 \subsection{Stability of the  equilibrium $E_4$}   \label{stability_E_4}

 We now analyze the stability of the equilibrium $E_4 = (U, 0, 1, C(q-\ell), W, 0),$
 which is the only feasible equilibrium in $\bar{\mathcal{U}}$ in absence of weeds, i.e. for $h_1 > a.$
 
 Based on the discussion raised by Remark  \ref{remark_2} in this case it is sufficient to
 analyze the stability of the equilibrium $E_4$ with respect to the only components
 $(F_4 = 0, X_4 = 1, N_4 = C(q - \ell), M_4 = 0).$
  
 In this case by denoting  $ \widetilde{{\bf Z}}(t) := (F (t), X(t), N(t), M(t))^T ,$ we may limit our
 analysis to the following system

 \begin{equation}  \label{91} 
 \displaystyle \frac{d}{d\,t} \widetilde{{\bf Z}}(t) = \widetilde{{\bf f}}( \widetilde{{\bf Z}}(t)),
 \end{equation} 
 for


 \begin{equation}
 \label{eq_92}
 \widetilde{{\bf f}}(\widetilde{{\bf Z}})=
 \left(
 \begin{array}{c}
 \displaystyle -rF\left[\left(\frac{n}{r}-M\right)+\chi\right] \vspace{0.2cm}\\
 \displaystyle -A\left(X+\frac{\alpha}{A}\right)(X-X_4) \vspace{0.2cm}\\
 \displaystyle -\frac{N}{C}(N-N_4) \vspace{0.2cm}\\
 \displaystyle -M\left[(h_1-a)+\frac{M}{C_2}+\delta(1-W)\right] 
 \end{array}
 \right), \quad \widetilde{{\bf Z}}\in\widetilde{\mathcal{U}}.
 \end{equation}

 where  $\widetilde{\mathcal{U}}:= (0, \bar{F})  \times (0,1) \times (0, \bar{N}) \times  (0, \bar{M}).$

 We may remark that, due to the fact that
 
  \begin{equation}  \label{93} 
 \displaystyle \frac{d}{d\,t} M(t) = - M(t) \left[(h_1 - a)  +  \frac{M(t)}{C_2}   +  \delta (1- W(t))\right]  <0,
 \end{equation} 
 for any $M(t) > 0,$   there exists a $t^* > 0$  such that, for any $t > t^*, $  $ M(t) < \displaystyle \frac{n}{r}.$

Since we are going to analyze the asymptotic behavior of System (\ref{91} ) we
 may take this into account.
 
  In order to analyze the stability of the equilibrium $E_4,$  we take as Lyapunov
 function

   \begin{equation}  \label{94} 
 \displaystyle V(\widetilde{{\bf Z}}):= \frac{1}{2}      \left[ \beta_F F^2  +  \alpha_X (X-1)^2  +  \beta_N (N- N_4)^2 + \beta_M M^2 \right]              ,
 \end{equation} 
 
 where  $\beta_F, \alpha_X,  \beta_N,  \beta_M $  are  positive constants to be suitably chosen.
 
As a consequence of the definition, the function $V  \in C^1(\widetilde{\mathcal{U}})$ and

   \begin{equation}  \label{95} 
  V(\widetilde{{\bf Z}}) \ge 0 \quad \textnormal{for all} \quad \widetilde{{\bf Z}} \in  \bar{\widetilde{\mathcal{U}}},
 \end{equation} 
 
 and 
 
 \begin{equation}  \label{96} 
 V(\widetilde{{\bf Z}}) = 0 \quad \textnormal{iff} \quad  V(\widetilde{{\bf Z}})  =  (0, 1, N_4,0)^T.
 \end{equation} 
 
 Moreover the derivative   of  $V$ along the trajectories of System (\ref{91}) is given by

 
 {\small
\begin{equation}
\label{eq_97}
\begin{array}{lll}
DV({\widetilde{\bf Z}}(t))  	& :=    & \displaystyle \mbox{grad} V(\widetilde{{\bf Z}}(t))\cdot\frac{d}{dt}\widetilde{{\bf Z}}(t) \vspace{0.2cm}\\
& =     & \displaystyle -\beta_F r (F(t))^2\left[\frac{n}{r}-M(t)+\chi F(t)\right] \vspace{0.2cm}\\
&       & \displaystyle -\left[\alpha_X A \left(X(t)+\frac{\alpha}{A}\right)(X(t)-1)^2 + \beta_N \frac{N(t)}{C}(N(t)-N_4)^2\right] \vspace{0.2cm}\\
&       & \displaystyle -\beta_M (M(t))^2 \left[\frac{M(t)}{C_2} + \delta(1-W(t))\right].
\end{array}
\end{equation} 
}
 
 It is then clear that

    \begin{equation}  \label{95} 
D V(\widetilde{{\bf Z}}) \le 0 \quad \textnormal{for all} \quad \widetilde{{\bf Z}} \in  \widetilde{\mathcal{U}}
 \end{equation} 
 
 and 
 
 \begin{equation}  \label{96} 
D V(\widetilde{{\bf Z}}) = 0 \quad \textnormal{iff} \quad  V(\widetilde{{\bf Z}})  =  (0, 1, N_4,0)^T.
 \end{equation} 
 
This leads to the following

\begin{theorem}      \label{Theorem_4}
 Under the assumptions $a > 0$ and $h_1 > a,$  the equilibrium $E_4$  is
 globally asymptotically stable in $\bar{\mathcal{U}}\setminus \{N=0  \}.  $
\end{theorem}

\section{Numerical experiments}     \label{numerics}

The numerical tests have been performed by solving system (\ref{adults+trees+plants}) using
the {\bf ode23s} Matlab built-in function. We consider the following ten different cases, depending on the choice
of the parameters $\lambda$, $\eta$, $\chi$, $h_1$. The values of the other parameters are given in Table  \ref{tab:param}. 
For each case, we report the plot of time evolution of the six state variables and a table with initial conditions and the equilibrium
reached at the final time t=100.

\begin{itemize}
	\item Case $E_1$: $\lambda=0$ , $\eta=0$, $\chi=0.001$, $h_1=0$ (Fig. \ref{fig_E_1}, Table \ref{tab_E_1});
	\item Case $E_1^{'}$: $\lambda=0$ , $\eta=0$, $\chi=0.01$, $h_1=0$ (Fig. \ref{fig_E_1_bis}, Table \ref{tab_E_1_bis});
	\item Case $E_2$: $\lambda=0.5$ , $\eta=0.1$, $\chi=0.001$, $h_1=0$ (Fig. \ref{fig_E_2}, Table \ref{tab_E_2});
	\item Case $E_2^{'}$: $\lambda=0.5$ , $\eta=0.1$, $\chi=0.01$, $h_1=0$ (Fig. \ref{fig_E_2_bis}, Table \ref{tab_E_2_bis});
	\item Case $E_3$: $\lambda=0.8$ , $\eta=0.1$, $\chi=0.001$, $h_1=0$ (Fig. \ref{fig_E_3}, Table \ref{tab_E_3});
	\item Case $E_3^{'}$: $\lambda=0.8$ , $\eta=0.1$, $\chi=0.01$, $h_1=0$ (Fig. \ref{fig_E_3_bis}, Table \ref{tab_E_3_bis});
	\item Case $E_4$: $\lambda=0.8$ , $\eta=0.1$, $\chi=0.001$, $h_1=0.5$ (Fig. \ref{fig_E_4}, Table \ref{tab_E_4});
	\item Case $E_4^{'}$: $\lambda=0.8$ , $\eta=0.1$, $\chi=0.01$, $h_1=0.5$ (Fig. \ref{fig_E_4_bis}, Table \ref{tab_E_4_bis});
	\item Case $\hat{E}_4$: $\lambda=0.5$ , $\eta=0.1$, $\chi=0.001$, $h_1=0.5$ (Fig. \ref{fig_E_4_tris}, Table \ref{tab_E_4_tris});
	\item Case $\hat{E}_4^{'}$: $\lambda=0.5$ , $\eta=0.1$, $\chi=0.01$, $h_1=0.5$ (Fig. \ref{fig_E_4_quadris}, Table \ref{tab_E_4_quadris}).
\end{itemize}

\subsection{Cases $E_1$ and $E_1^{'}$}

In the first scenario, we set the trees and weeds infection rates ($\lambda$ and $\eta$) and the weeds eradication parameter ($h_1$) to zero. As expected, at equilibrium, the fraction of healthy trees approaches the value $X=1$, meaning that all olive trees are healthy, thus the epidemic dies down.
This behavior occurs irrespective of the values assumed
by the insect intraspecific competition rate ($\chi=0.001$ in case $E_1$ and $\chi=0.01$ in case $E_1^{'}$).

\subsection{Cases $E_2$ and $E_2^{'}$}

In the second scenario, we set the trees and weeds infection rates ($\lambda$ and $\eta$) to 0.5 and 0.1, respectively, and the weeds eradication parameter ($h_1$) to zero. 
When the insect intraspecific competition rate is low ($\chi=0.001$, case $E_2$), at equilibrium, 75\% of trees is healthy, 
meaning that the
epidemic has not expired.
On the other hand, setting the insect intraspecific competition rate to a higher value ($\chi=0.01$, case $E_2^{'}$), 
at equilibrium, the fraction of healthy trees approaches the value $X=1$, thus the epidemic dies down.

\subsection{Cases $E_3$ and $E_3^{'}$}

In the third scenario, we set the trees and weeds infection rates ($\lambda$ and $\eta$) to 0.8 and 0.1, respectively, and the weeds eradication parameter ($h_1$) to zero.
When the insect intraspecific competition rate is low ($\chi=0.001$, case $E_3$), at equilibrium, the total population of trees $N$ has expired.
Increasing instead the insect intraspecific competition rate ($\chi=0.01$, case $E_3^{'}$),
at equilibrium, the fraction of healthy trees approaches again the value $X=1$, thus the epidemic dies down.

\subsection{Cases $E_4$, $E_4^{'}$, $\hat{E}_4$, $\hat{E}_4^{'}$}

In the last scenario, we first set the trees and weeds infection rates ($\lambda$ and $\eta$) to 0.8 and 0.1, respectively, and the weeds eradication parameter ($h_1$) to 0.5. With a low insect intraspecific competition rate ($\chi=0.001$, case $E_4$), the total population of trees $N$ initially reduces considerably, but after $t=40$, due to the effectiveness of weeds eradication, it starts to increase. At equilibrium, 100\% of trees is healthy ($X=1$),
thus the
epidemic dies down. 
With a larger insect intraspecific competition rate ($\chi=0.01$, case $E_4^{'}$), the fraction of healthy trees approaches the equilibrium value $X=1$ earlier.

Then, in cases $\hat{E}_4$ and $\hat{E}_4^{'}$, we set the trees and weeds infection rates ($\lambda$ and $\eta$) to 0.5 and 0.1, respectively, and the weeds eradication parameter ($h_1$) to 0.5. The dynamics is analogous to the previous cases $E_4$, $E_4^{'}$, 
since,  at equilibrium, 100\% of trees is healthy ($X=1$) and the epidemic expires. 

\pagebreak

\begin{figure}
	\begin{center}
		\includegraphics[width=16cm]{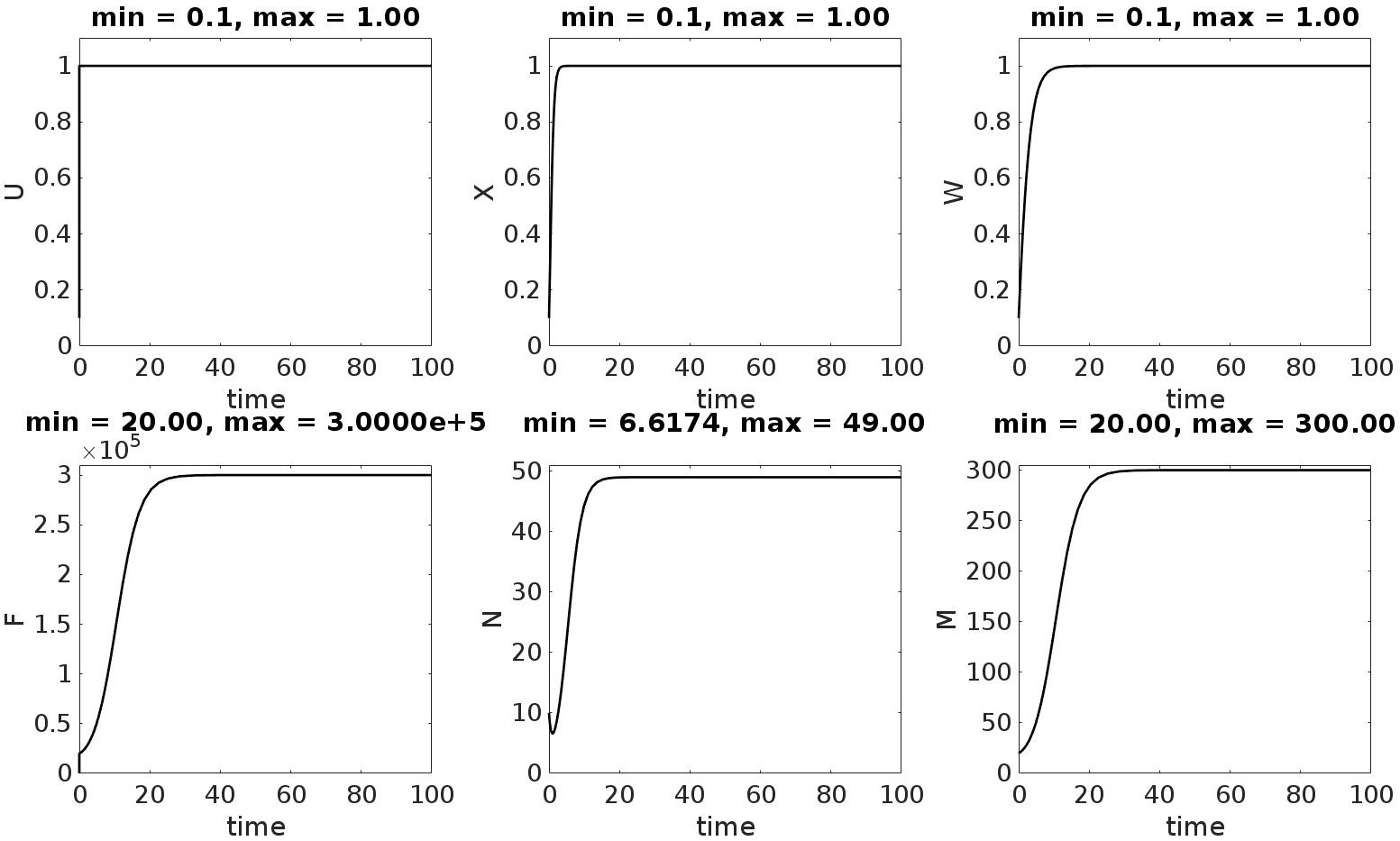}
		\caption{Case $E_1$. Plots of time evolution of the state variables.}
		\label{fig_E_1}
	\end{center}
\end{figure}

\begin{table}
	\begin{center}
		\begin{tabular}{c|c|c}
			\hline
			Variable	& t=0			& t=100 \\
			\hline
			U		& $ 1.0000e-01 $  	& $ 1.0000e+00 $ \\
			F		& $ 2.0000e+01 $  	& $ 3.0000e+05 $ \\
			X		& $ 1.0000e-01 $  	& $ 1.0000e+00 $ \\
			N		& $ 1.0000e+01 $  	& $ 4.9000e+01 $ \\
			W		& $ 1.0000e-01 $  	& $ 1.0000e+00 $ \\
			M		& $ 2.0000e+01 $  	& $ 3.0000e+02 $ \\
			\hline
		\end{tabular}
		\caption{Case $E_1$. Initial (t=0) and final (t=100) configuration of the state variables.}
		\label{tab_E_1}
	\end{center}
\end{table}

\begin{figure}
	\begin{center}
		\includegraphics[width=16cm]{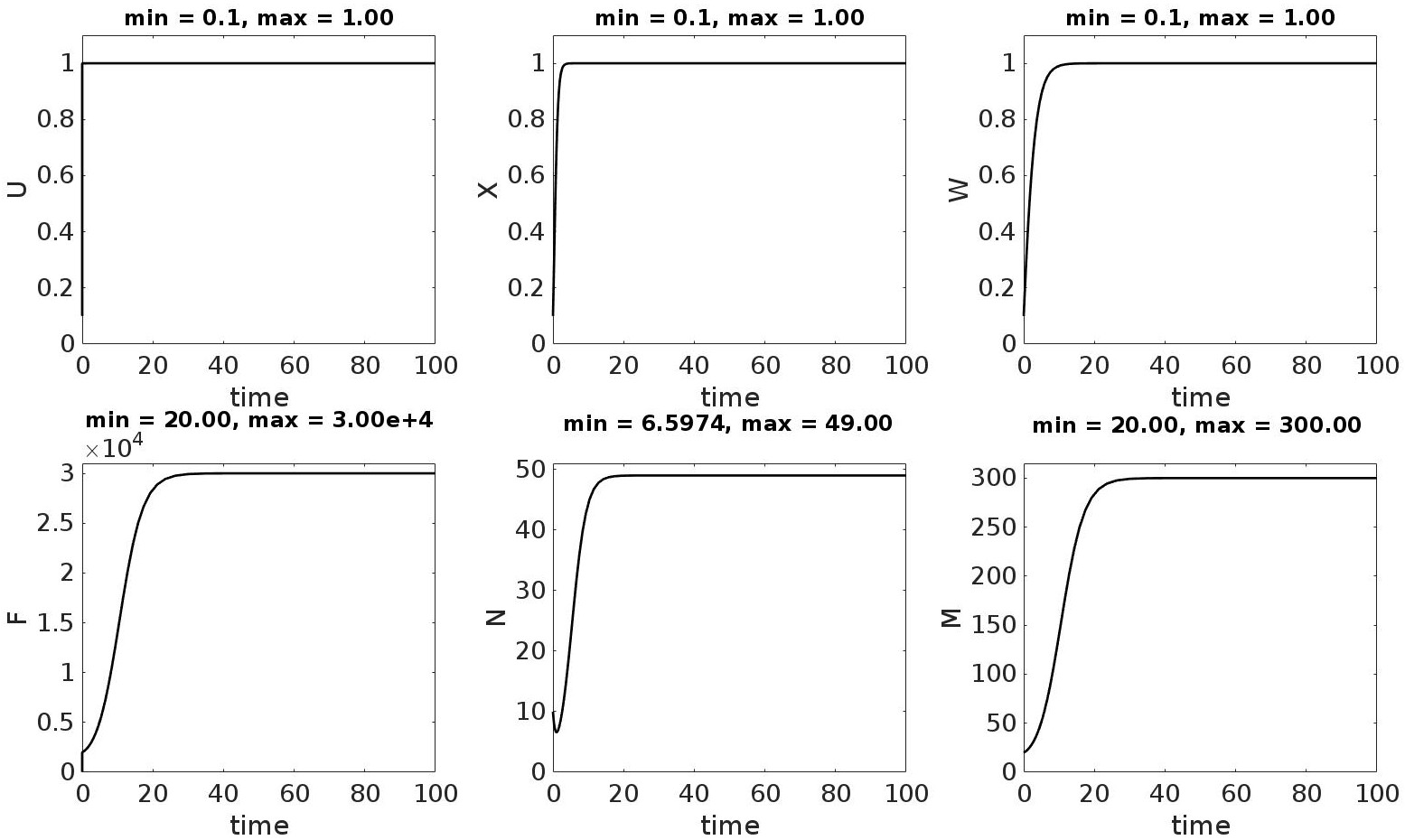}
		\caption{Case $E_1^{'}$. Plots of time evolution of the state variables.}
		\label{fig_E_1_bis}
	\end{center}
\end{figure}

\begin{table}
	\begin{center}
		\begin{tabular}{c|c|c}
			\hline
			Variable        & t=0                   & t=100 \\
			\hline
			U   		& $ 1.0000e-01 $   	& $ 1.0000e+00 $ \\
			F   		& $ 2.0000e+01 $   	& $ 3.0000e+04 $ \\
			X   		& $ 1.0000e-01 $   	& $ 1.0000e+00 $ \\
			N   		& $ 1.0000e+01 $   	& $ 4.9000e+01 $ \\
			W   		& $ 1.0000e-01 $   	& $ 1.0000e+00 $ \\
			M   		& $ 2.0000e+01 $   	& $ 3.0000e+02 $ \\
			\hline
		\end{tabular}
		\caption{Case $E_1^{'}$. Initial (t=0) and final (t=100) configuration of the state variables.}
		\label{tab_E_1_bis}
	\end{center}
\end{table}

\begin{figure}
	\begin{center}
		\includegraphics[width=16cm]{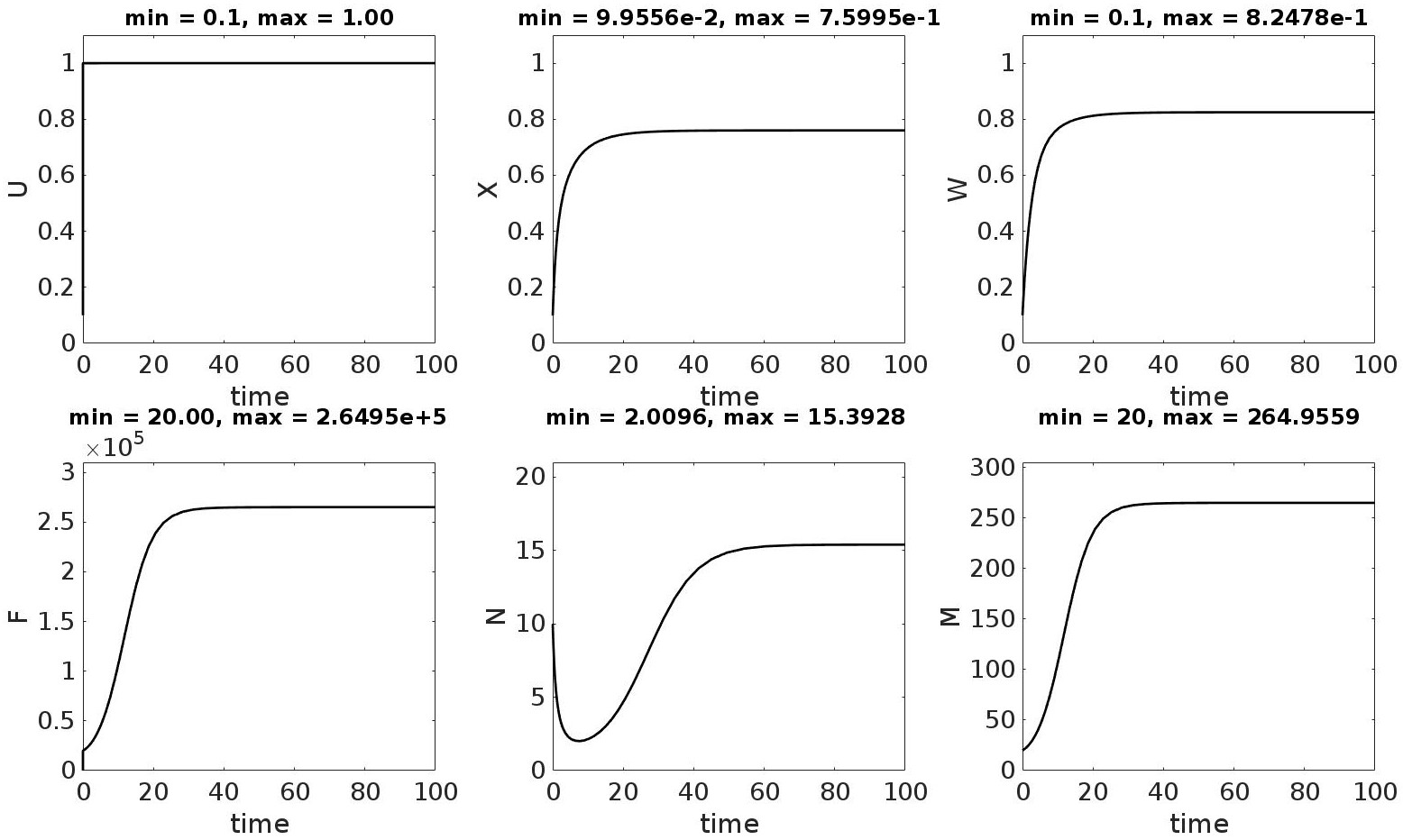}
		\caption{Case $E_2$. Plots of time evolution of the state variables.}
		\label{fig_E_2}
	\end{center}
\end{figure}

\begin{table}
	\begin{center}
		\begin{tabular}{c|c|c}
			\hline
			Variable	& t=0          		& t=100 \\
			\hline
			U   		& $ 1.0000e-01 $ 	& $ 1.0000e+00 $ \\
			F   		& $ 2.0000e+01 $  	& $ 2.6495e+05 $ \\
			X   		& $ 1.0000e-01 $  	& $ 7.5995e-01 $ \\
			N   		& $ 1.0000e+01 $  	& $ 1.5393e+01 $ \\
			W   		& $ 1.0000e-01 $  	& $ 8.2478e-01 $ \\
			M   		& $ 2.0000e+01 $  	& $ 2.6496e+02 $ \\
			\hline
		\end{tabular}
		\caption{Case $E_2$. Initial (t=0) and final (t=100) configuration of the state variables.}
		\label{tab_E_2}
	\end{center}
\end{table}

\begin{figure}
	\begin{center}
		\includegraphics[width=16cm]{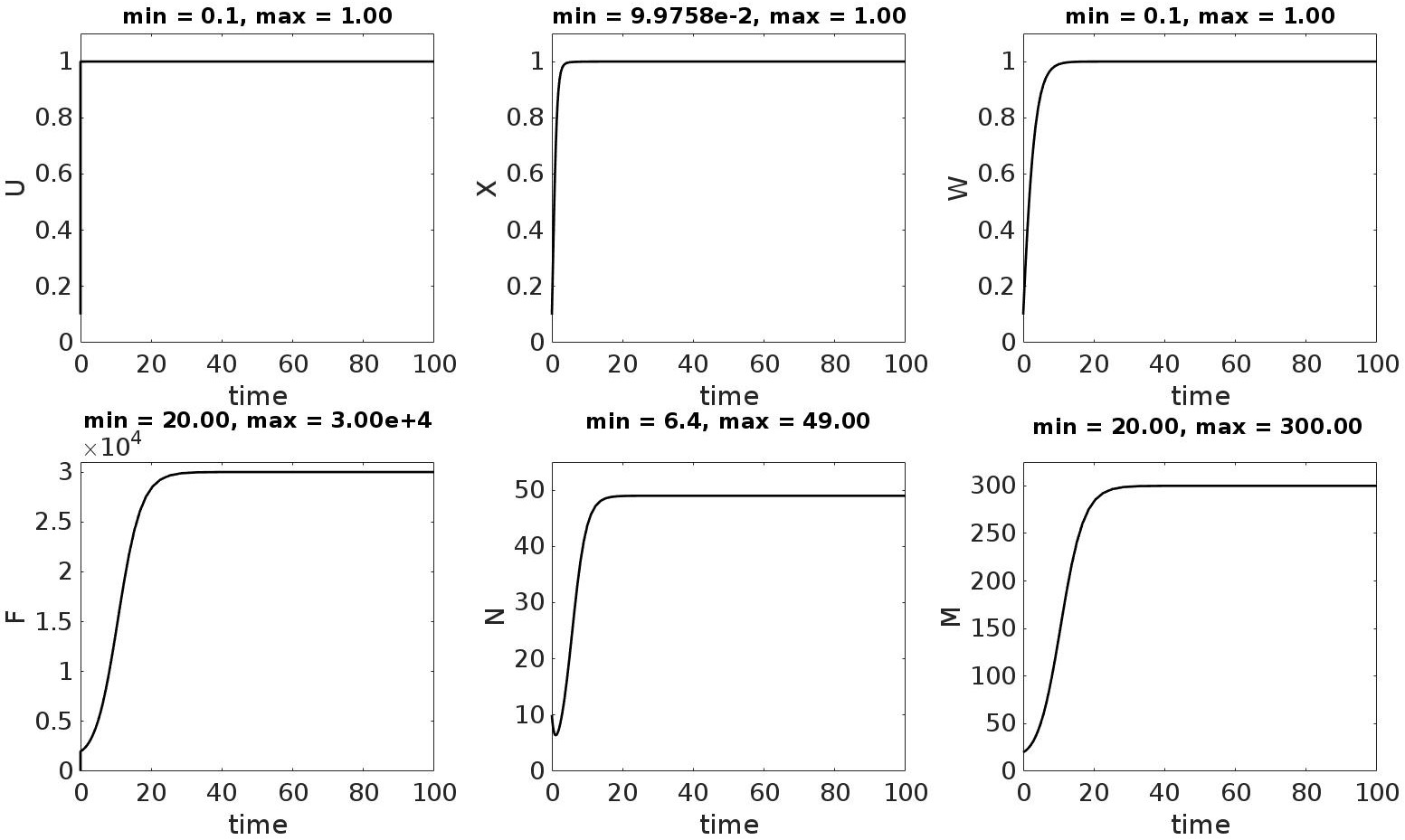}
		\caption{Case $E_2^{'}$. Plots of time evolution of the state variables.}
		\label{fig_E_2_bis}
	\end{center}
\end{figure}

\begin{table}
	\begin{center}
		\begin{tabular}{c|c|c}
			\hline
			Variable        & t=0                   & t=100 \\
			\hline
			U   		& $ 1.0000e-01 $   	& $ 1.0000e+00 $ \\
			F   		& $ 2.0000e+01 $   	& $ 3.0000e+04 $ \\
			X   		& $ 1.0000e-01 $   	& $ 1.0000e+00 $ \\
			N   		& $ 1.0000e+01 $   	& $ 4.9000e+01 $ \\
			W   		& $ 1.0000e-01 $   	& $ 1.0000e+00 $ \\
			M   		& $ 2.0000e+01 $   	& $ 3.0000e+02 $ \\
			\hline
		\end{tabular}
		\caption{Case $E_2^{'}$. Initial (t=0) and final (t=100) configuration of the state variables.}
		\label{tab_E_2_bis}
	\end{center}
\end{table}

\begin{figure}
	\begin{center}
		\includegraphics[width=16cm]{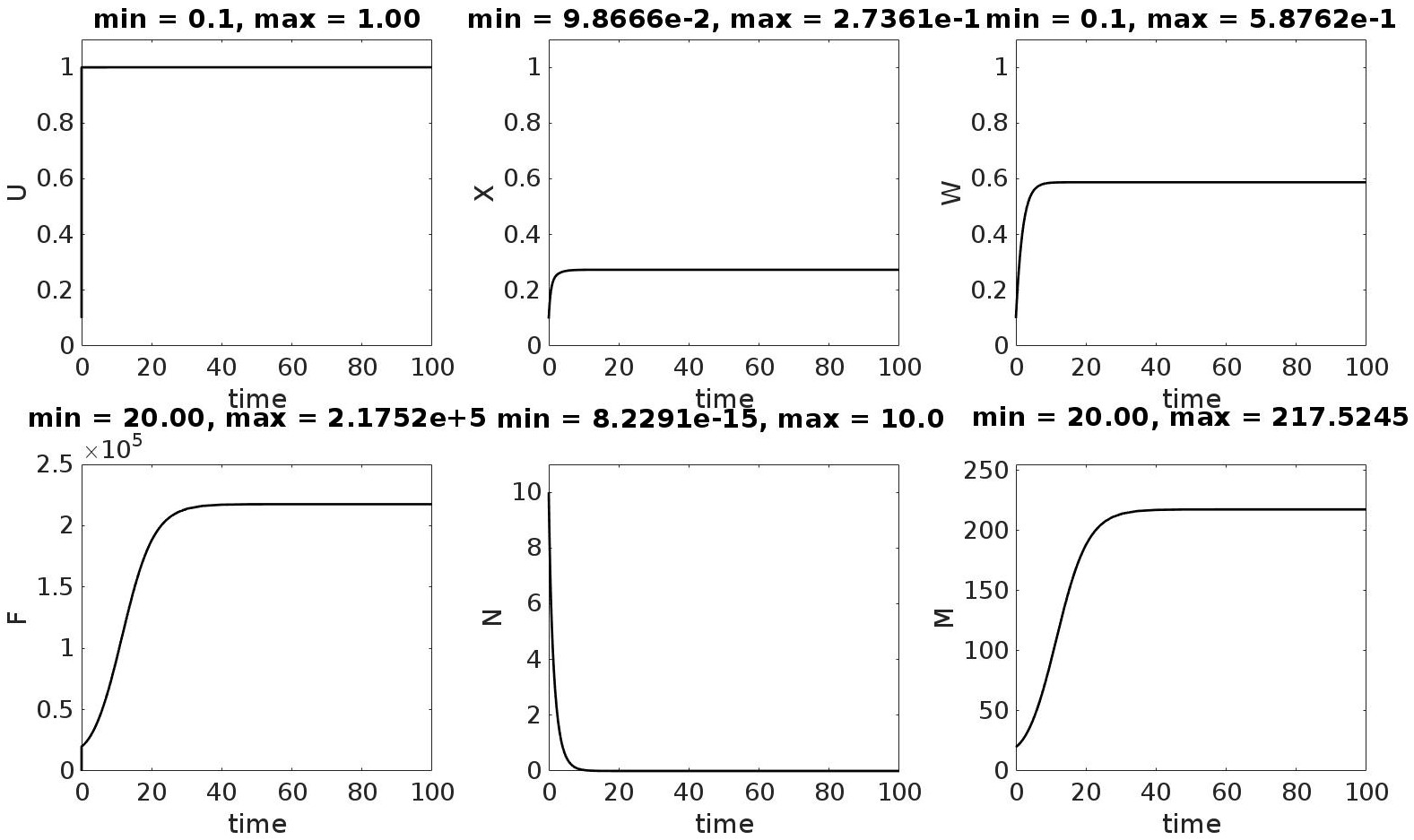}
		\caption{Case $E_3$. Plots of time evolution of the state variables.}
		\label{fig_E_3}
	\end{center}
\end{figure}

\begin{table}
	\begin{center}
		\begin{tabular}{c|c|c}
			\hline
			Variable	& t=0          		& t=100 \\
			\hline
			U   		& $ 1.0000e-01 $   	& $ 9.9999e-01 $ \\
			F   		& $ 2.0000e+01 $   	& $ 2.1752e+05 $ \\ 
			X   		& $ 1.0000e-01 $   	& $ 2.7360e-01 $ \\
			N   		& $ 1.0000e+01 $   	& $ 8.2291e-15 $ \\
			W   		& $ 1.0000e-01 $   	& $ 5.8762e-01 $ \\
			M   		& $ 2.0000e+01 $   	& $ 2.1752e+02 $ \\
			\hline
		\end{tabular}
		\caption{Case $E_3$. Initial (t=0) and final (t=100) configuration of the state variables.}
		\label{tab_E_3}
	\end{center}
\end{table}

\begin{figure}
	\begin{center}
		\includegraphics[width=16cm]{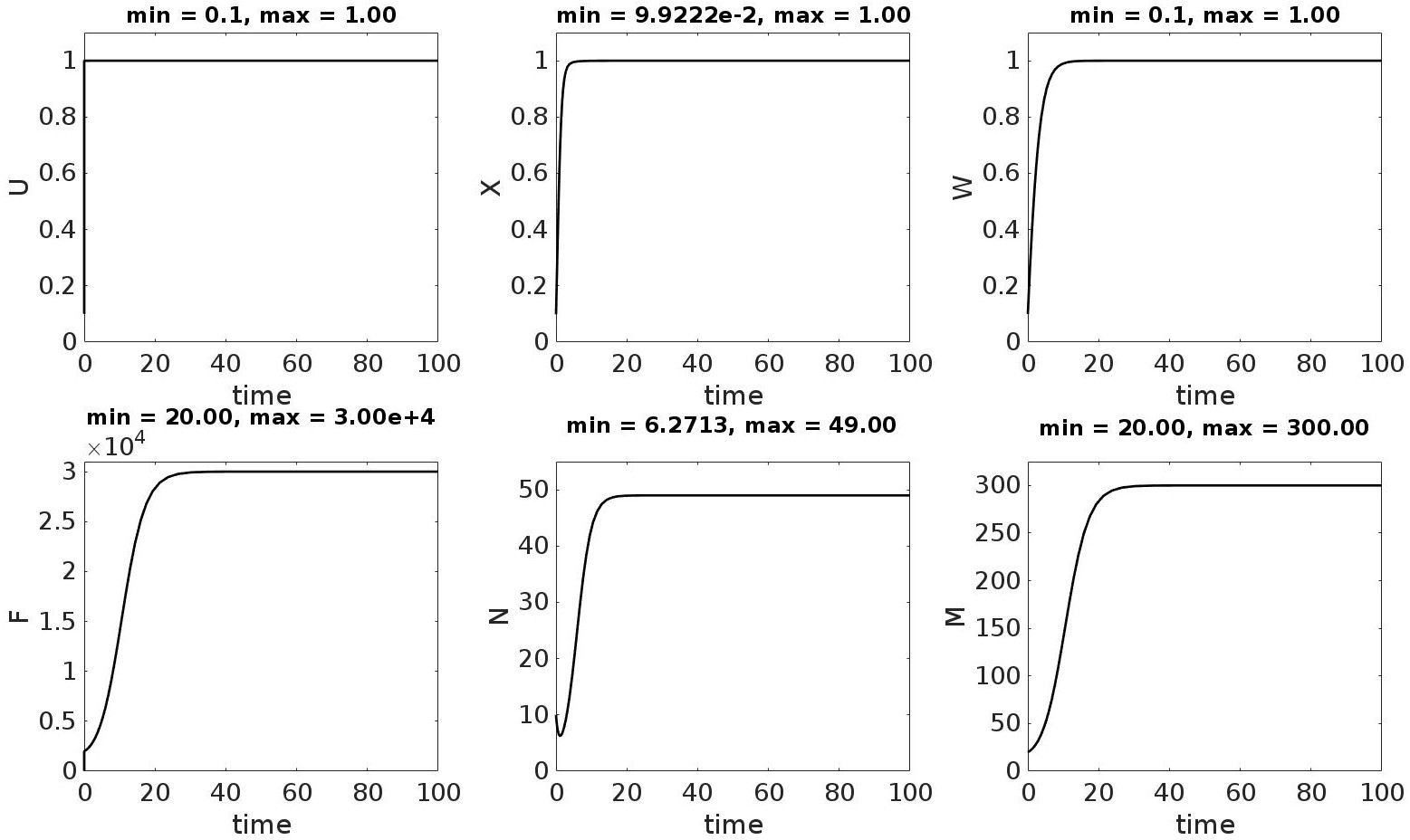}
		\caption{Case $E_3^{'}$. Plots of time evolution of the state variables.}
		\label{fig_E_3_bis}
	\end{center}
\end{figure}

\begin{table}
	\begin{center}
		\begin{tabular}{c|c|c}
			\hline
			Variable        & t=0                   & t=100 \\
			\hline
			U   		& $ 1.0000e-01 $  	& $ 1.0000e+00 $ \\
			F   		& $ 2.0000e+01 $   	& $ 3.0000e+04 $ \\
			X   		& $ 1.0000e-01 $   	& $ 1.0000e+00 $ \\
			N   		& $ 1.0000e+01 $   	& $ 4.9000e+01 $ \\
			W   		& $ 1.0000e-01 $  	& $ 1.0000e+00 $ \\
			M   		& $ 2.0000e+01 $   	& $ 3.0000e+02 $ \\
			\hline
		\end{tabular}
		\caption{Case $E_3^{'}$. Initial (t=0) and final (t=100) configuration of the state variables.}
		\label{tab_E_3_bis}
	\end{center}
\end{table}

\begin{figure}
	\begin{center}
		\includegraphics[width=16cm]{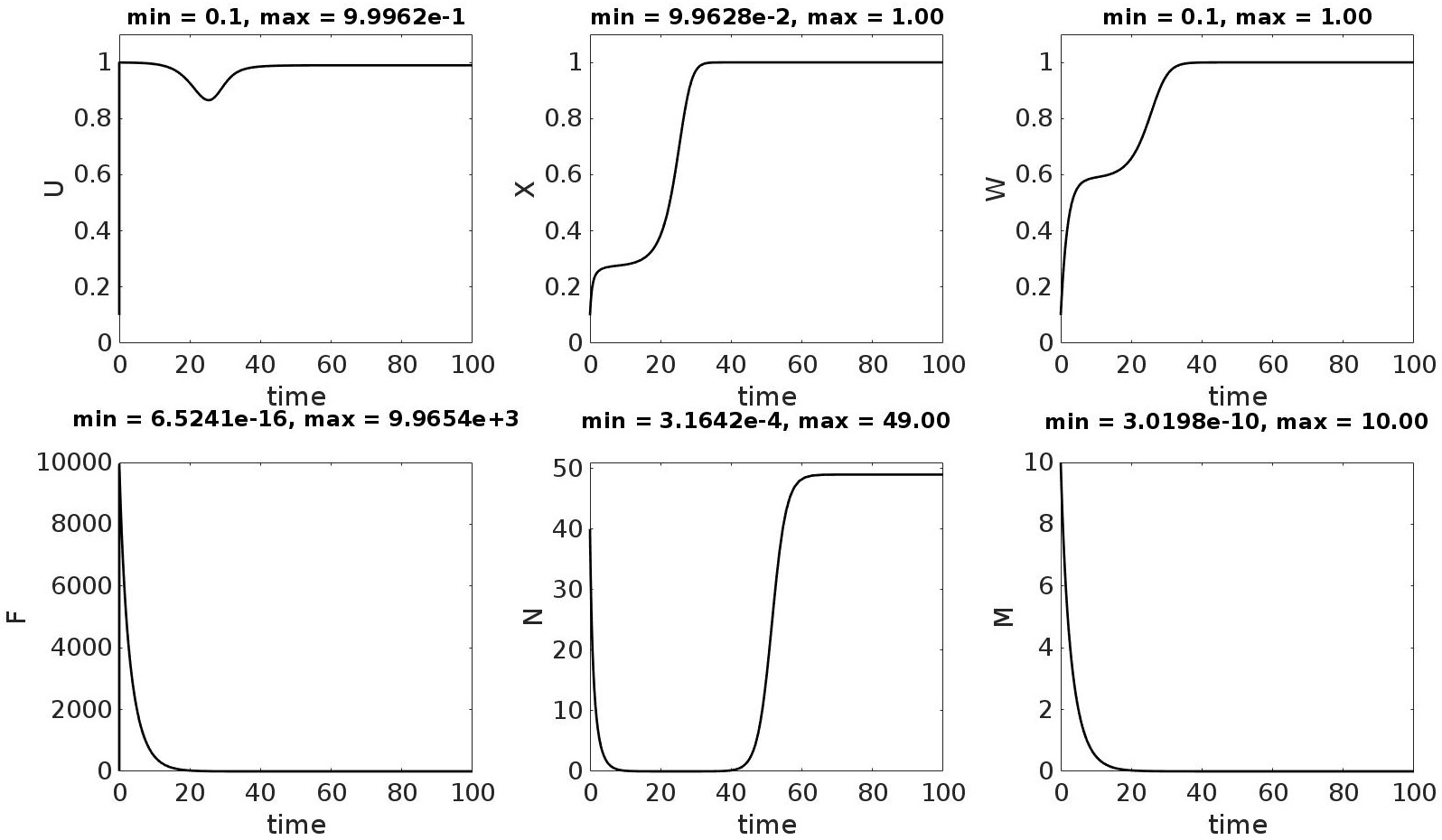}
		\caption{Case $E_4$. Plots of time evolution of the state variables.}
		\label{fig_E_4}
	\end{center}
\end{figure}

\begin{table}
	\begin{center}
		\begin{tabular}{c|c|c}
			\hline
			Variable	& t=0          		& t=100 \\
			\hline
			U   		& $ 1.0000e-01 $ 	& $ 9.8974e-01 $ \\
			F   		& $ 1.0000e+01 $   	& $ 6.5241e-16 $ \\ 
			X   		& $ 1.0000e-01 $   	& $ 1.0000e+00 $ \\
			N   		& $ 4.0000e+01 $   	& $ 4.9000e+01 $ \\
			W   		& $ 1.0000e-01 $   	& $ 1.0000e+00 $ \\
			M   		& $ 1.0000e+01 $   	& $ 3.0198e-10 $ \\
			\hline
		\end{tabular}
		\caption{Case $E_4$. Initial (t=0) and final (t=100) configuration of the state variables.}
		\label{tab_E_4}
	\end{center}
\end{table}

\begin{figure}
	\begin{center}
		\includegraphics[width=16cm]{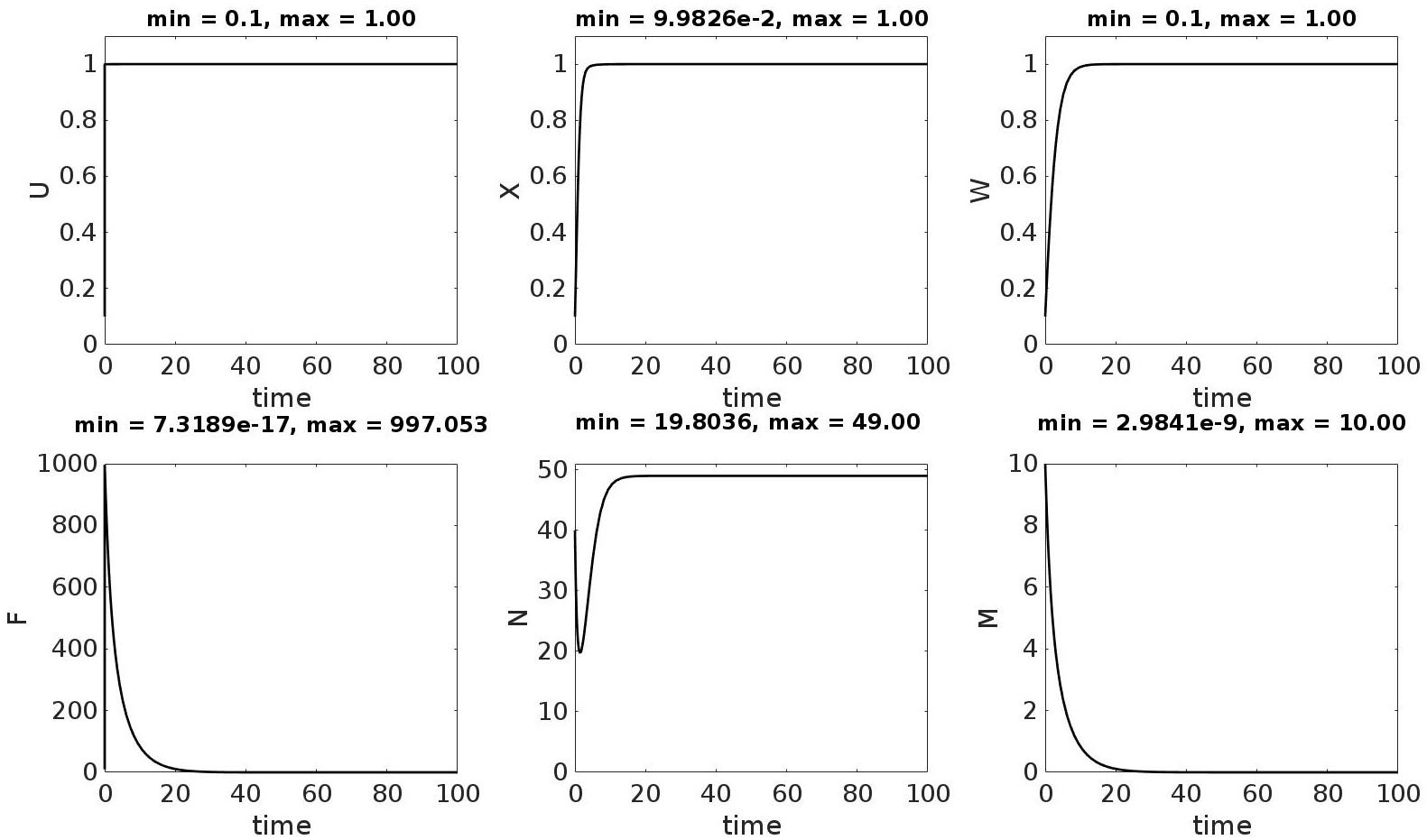}
		\caption{Case $E_4^{'}$. Plots of time evolution of the state variables.}
		\label{fig_E_4_bis}
	\end{center}
\end{figure}

\begin{table}
	\begin{center}
		\begin{tabular}{c|c|c}
			\hline
			Variable        & t=0                   & t=100 \\
			\hline
			U   		& $ 1.0000e-01 $   	& $ 1.0000e+00 $ \\
			F   		& $ 1.0000e+01 $   	& $ 7.3189e-17 $ \\
			X   		& $ 1.0000e-01 $   	& $ 1.0000e+00 $ \\
			N   		& $ 4.0000e+01 $   	& $ 4.9000e+01 $ \\
			W   		& $ 1.0000e-01 $   	& $ 1.0000e+00 $ \\
			M   		& $ 1.0000e+01 $   	& $ 2.9841e-09 $ \\
			\hline
		\end{tabular}
		\caption{Case $E_4^{'}$. Initial (t=0) and final (t=100) configuration of the state variables.}
		\label{tab_E_4_bis}
	\end{center}
\end{table}

\begin{figure}
	\begin{center}
		\includegraphics[width=16cm]{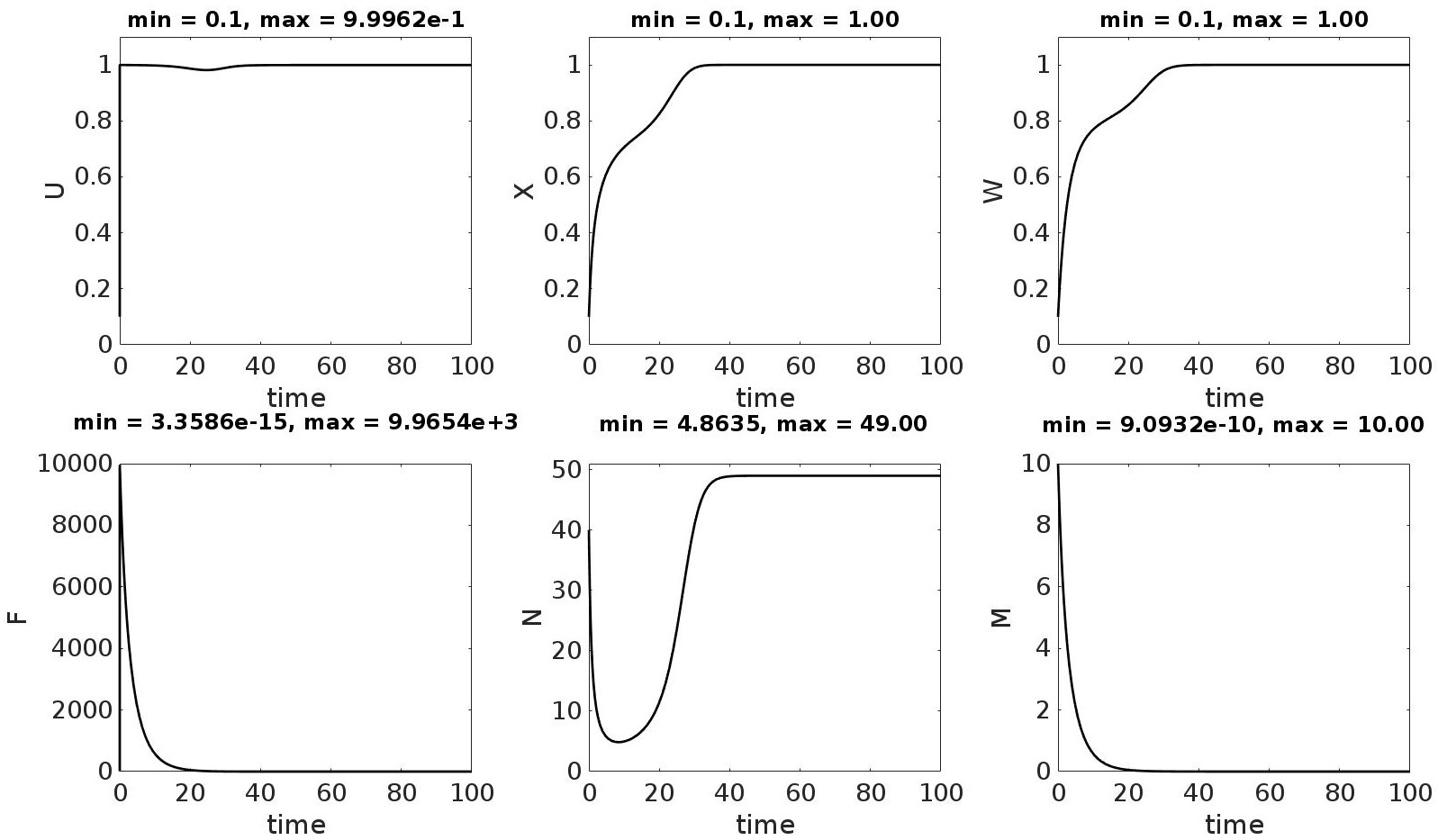}
		\caption{Case $\hat{E}_4$. Plots of time evolution of the state variables.}
		\label{fig_E_4_tris}
	\end{center}
\end{figure}

\begin{table}
	\begin{center}
		\begin{tabular}{c|c|c}
			\hline
			Variable        & t=0                   & t=100 \\
			\hline
			U               & $ 1.0000e-01 $        & $ 9.9950e-01 $ \\
			F               & $ 1.0000e+01 $        & $ 3.3586e-15 $ \\
			X               & $ 1.0000e-01 $        & $ 1.0000e+00 $ \\
			N               & $ 4.0000e+01 $        & $ 4.9000e+01 $ \\
			W               & $ 1.0000e-01 $        & $ 1.0000e+00 $ \\
			M               & $ 1.0000e+01 $        & $ 9.0932e-10 $ \\
			\hline
		\end{tabular}
		\caption{Case $\hat{E}_4$. Initial (t=0) and final (t=100) configuration of the state variables.}
		\label{tab_E_4_tris}
	\end{center}
\end{table}

\begin{figure}
	\begin{center}
		\includegraphics[width=16cm]{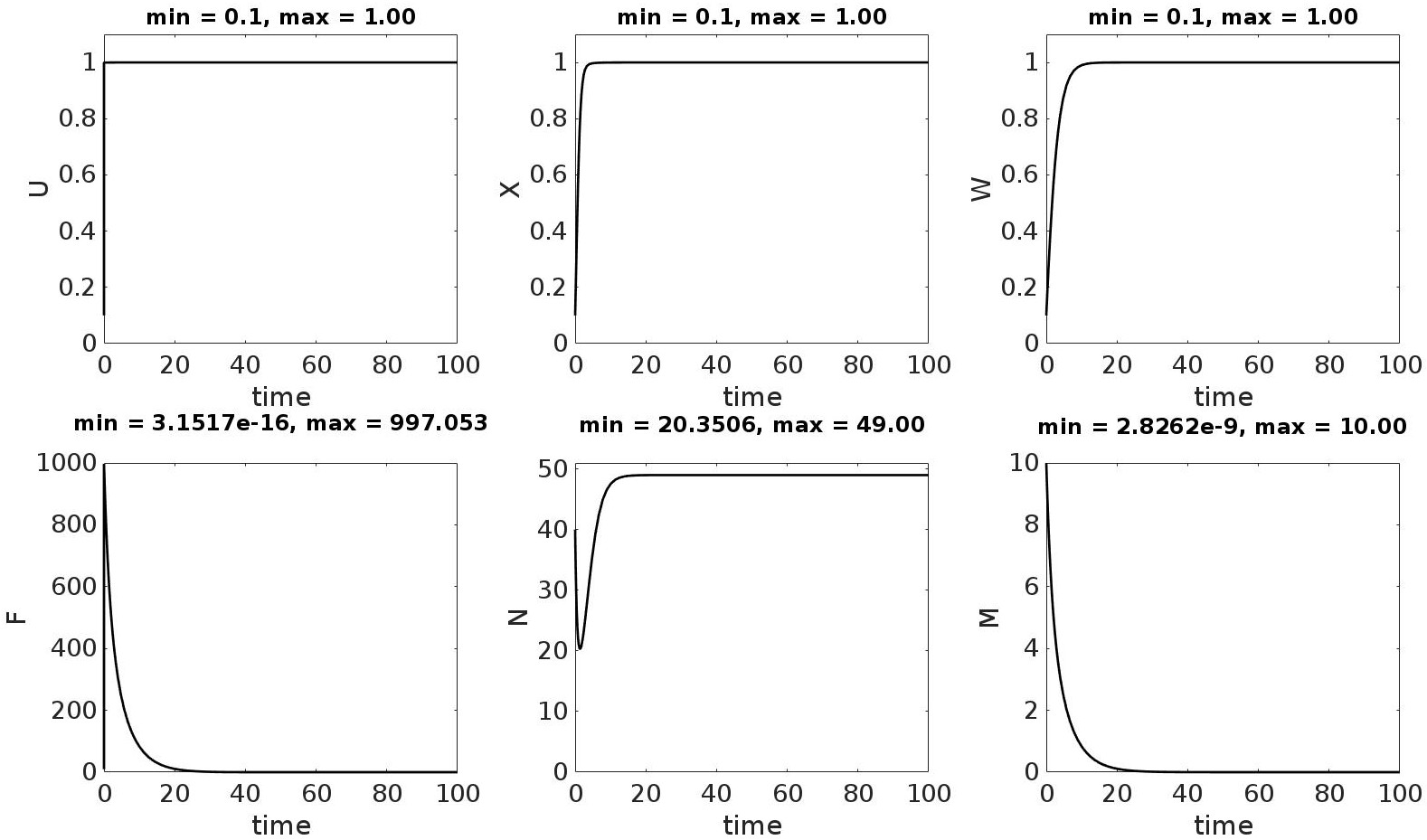}
		\caption{Case $\hat{E}_4^{'}$. Plots of time evolution of the state variables.}
		\label{fig_E_4_quadris}
	\end{center}
\end{figure}

\begin{table}
	\begin{center}
		\begin{tabular}{c|c|c}
			\hline
			Variable        & t=0                   & t=100 \\
			\hline
			U   		& $ 1.0000e-01 $   	& $ 1.0000e+00 $ \\
			F   		& $ 1.0000e+01 $   	& $ 3.1517e-16 $ \\
			X   		& $ 1.0000e-01 $   	& $ 1.0000e+00 $ \\
			N   		& $ 4.0000e+01 $   	& $ 4.9000e+01 $ \\
			W   		& $ 1.0000e-01 $   	& $ 1.0000e+00 $ \\
			M   		& $ 1.0000e+01 $   	& $ 2.8262e-09 $ \\
			\hline
		\end{tabular}
		\caption{Case $\hat{E}_4^{'}$. Initial (t=0) and final (t=100) configuration of the state variables.}
		\label{tab_E_4_quadris}
	\end{center}
\end{table}

\section{Concluding remarks}   \label{Concluding remarks}

 The above analysis has shown that four main cases are feasible for the equilibria 
 of the dynamical system describing the evolution of   {\it Xylella fastidiosa}  epidemics within olive orchard agroecosystems.  The equilibrium $E_1$ corresponds to a disease free
 ecosystem (see Figure \ref{fig_E_1}); equilibrium  $E_2$  corresponds to coexistence of a non-
 trivial olive tree biomass and infective insects, which we may conjecture (also
 supported by the numerical experiments) is possible only for a sufficiently small
 value of $\lambda,$  i.e. for more resistant olive tree cultivars (see Figure \ref{fig_E_2}); the equilibrium
  $E_2$ may degenerate into  $E_3$ for a sufficiently large value of $\lambda,$ 
  i.e. for less resistant olive tree cultivars, which leads to the complete disappearance of the olive tree
 biomass (see Figure \ref{fig_E_3}). Finally the equilibria   $E_4$ and $\widehat{E_4}$ correspond to eradication of
 the insect population, by a significant reduction of the weed reproduction rate
 (see Figures \ref{fig_E_4} and \ref{fig_E_4_tris}).
 
The  outcomes  of  the numerical simulations, reported in Figures \ref{fig_E_1_bis}, \ref{fig_E_2_bis}, \ref{fig_E_3_bis}, \ref{fig_E_4_bis}, \ref{fig_E_4_quadris},  
 elucidate the  crucial role of
 the parameter $\chi.$   From Equation
 (\ref{eq_2}) we may recognize the role of $\displaystyle
 \frac{M}{\chi}$  as   the effective carrying capacity of   the
 insect population $F;$ for $ \chi_1  \le \chi_2$ we   have
 
 $$\displaystyle \frac{M}{\chi_1}  \ge   \displaystyle \frac{M}{\chi_2},  $$
 which implies,  for   $\chi_1,$  a  possible  larger  population
 of insects, hence a possible larger   infective  insect
 population, which may lead to a larger force of infection on olive
 trees, thus making  the   coexistence of insects and trees
 unlikely.

 Hence, once an epidemic has started, in absence of intervention our model
predicts a possible full collapse of the olive tree biomass, with a significant socio-economic impact.
About agronomic practices, the most interesting result concerns the role of
weeds for the eradication of a { \it X. fastidiosa} epidemic. Indeed, according to our model,
a sufficient reduction of the weed biomass may lead to a significant decay of the
 insect populations, and consequently to the eventual eradication of the epidemic.
 Weeds, also present in the relevant  olive orchard,  represent the main feeding resource of the insects at their juvenile stage.  We
 have to be aware that  insects juvenile feed on a large variety of both weeds and ornamental plants, so that  a particular attention has to be paid not only
 to the usual spontaneous plants emerging in the olive orchard itself, but also to
 any kind of ornamental plant existing in its near neighborhood \cite{white2017}.
 A second possible strategy for prevention and control of a  { \it X. fastidiosa}  epidemic,
 which has been confirmed by our analysis, concerns the substitution of the
 currently cultivated  olive tree cultivars by more resistant ones. This approach has already
 been suggested by the local agriculture Authorities in Southern Apulia. But
 we have to take into account that, different from the usual good agronomic
 practice of weed elimination, the substitution of a cultivar is both money and time
 much more expensive; it takes a long time before a new planted or grafted
 tree reaches the level of production of an existing one. On the other hand,
 it goes without saying that the impact of the cultivar on the quality of the
 extracted olive oil can be dramatic for the local economy. As an example,
 Apulia has a great international reputation for the production of a variety of
  olive oil of outstanding quality, based on the current cultivar all over the
region (see e.g. \cite{dugo2020}, https://bestoliveoils.org/brands/).



 Once again, we wish to conclude by warning the readers that validation
 of the model proposed here represents a key issue: although we have tried to
 make explicit the assumptions underlying our model, they have not yet been
 validated by comparison with experimental data. Therefore we caution that our
 results are far from being conclusive for {\it  X. fastidiosa} - {\it P. 
 spumarius} olive tree epidemics. However, it is desirable that with  further   experiments,  possibly driven by our models,  additional
 features are  added  that make them more realistic, so that  mathematical models  might provide the foundations
 for designing optimal control strategies by the  relevant public authorities.


\section*{Acknowledgments}
 The  authors  are indebted  to Professor Sebastian  Ani\c{t}a   of the University of Ia\c{s}i in Romania, and  Professor  
Ezio  Venturino of the University of  Turin for  relevant discussions on the mathematical   modelling aspects  of this   research  project.



\begin{thebibliography}{99}
	
	     	

	
	
	
	
    \bibitem{almeida2005} Almeida, R.P.P., Blua, M.J., Lopes, J.R.S.,
    Purcell, A.H., Vector transmission of {\it{Xylella    fastidiosa}}:
    applying fundamental knowledge to generate disease management
    strategies. Ann. Entomol. Soc. Am., 98 (2005), 775-786.

        

       

       
        
          \bibitem{anita_VK_scacchi_BMAB_2021} Ani\c{t}a, S., Capasso, V., Scacchi, S.,  Controlling the spatial  spread of a {\it Xylella} epidemic.  
        Bull. Math. Biology, (2021) 83:32.  doi: 
        10.1007/s11538-021-00861-z




 





    


  


    \bibitem{Boscia_2017} Boscia, D., Altamura, G.,  Saponari, M., Tavano, D.,  Zicca, S., Pollastro, P.,  Silletti, M.R.,  Savino,  V.N.,   Martelli, G.P.,
     Delle Donne,A. ,  Mazzotta,  S.,
    Signore, P.P., Troisi, M.,   Drazza, P.,   Conte, P.,  D'  Ostuni, V.,  Merico, S.,   Perrone, G.,   Specchia, F.,  Stanca, A.,  Tanieli, M.,
    Incidenza di Xylella in oliveti con disseccamento rapido. Informatore Agrario, 27(59-64) (2017), 47-50.




\bibitem{brunetti_VK_etal_ECOMOD_2020}  Brunetti, M., Capasso, V., Montagna, M., Venturino, E.,  A mathematical model for {\it Xylella fastidiosa}
epidemics in the Mediterranean regions. Promoting good agronomic practices for their effective control.
Ecol Model 432 (2020)109204. doi: 10.1016/j.ecolmodel.2020.109204	






    \bibitem{VK09}
    Capasso V.
    {\it Mathematical Structures of Epidemic Systems}, 2nd revised printing,
    Lecture Notes Biomath., Vol. ~97. Heidelberg: Springer-Verlag; 2009.










    \bibitem{carlucci2013} Carlucci, A., Lops, F., Marchi, G., Mugnai,
    L., Surico, G., Has {\it {\it  Xylella fastidiosa}} ``chosen'' olive
    trees to establish in the Mediterranean basin? Phytopathologia
    Mediterranea, 52 (2013), 541-544.

    
    \bibitem{cornara2017}  Cornara, D., et al., Transmission of {\it{Xylella    fastidiosa}} by naturally infected     {\it{ Philaenus spumarius}} (Hemiptera, Aphrophoridae) to different
    host plants. J. Appl. Entomol.  141 (2017), 80-87.

  

    \bibitem{dietz1982}  Dietz, K., Overall population patterns in the  transmission cycle of infectious disease agents. In
      {\it Population   Biology of Infectious Diseases}, R.M. Anderson, R.M. May, Editors.    Life Sciences Research Reports, Vol. ~25. Heidelberg: Springer-Verlag; 1982.

   
   
 \bibitem{dugo2020}   Dugo, L., Russo, M., Cacciola, F. et al.,  Determination of the Phenol and Tocopherol Content in Italian High-Quality Extra-Virgin Olive Oils by Using LC-MS and Multivariate Data Analysis. Food Anal. Methods 13 (2020), 1027–1041.
 doi: 10.1007/s12161-020-01721-7
   
   
     \bibitem{Fierro_2019}  Fierro, A., Liccardo, A.,  Porcelli, F. (2019). A lattice model to manage the vector and the infection of
     the {\it{Xylella    fastidiosa}} on olive trees. Scientific Reports, 9(1), 8723.

   
   

\bibitem{jeger_2018}	Jeger, M. et al. (EFSA PLH Panel, Updated pest categorisation of {\it Xylella fastidiosa}. EFSA Journal, 16(7)(2018), e05357.
doi: 	537 10.2903/j.efsa.2018.5357.

   
   
    \bibitem{martelli_2016} Martelli, G. P., Boscia, D., Porcelli, F., Saponari, M. , The olive quick decline syndrome in south-east Italy: a threatening phytosanitary emergency.
    European Journal of Plant Pathology, 144(2) (2016), 235-243.
    

    \bibitem{redak2004} Redak, R.A., Purcell, A.H.,  Lopes, J.R.S., Blua, M.J., Mizell, R.F. III, Andersen, P.C.,
    The  biology of xylem fluid-feeding insect vectors
    of {\it {\it  Xylella fastidiosa}} and their relation to disease epidemiology,
    applying fundamental knowledge to generate disease management.
    Annu. Rev. Entomol.  49 (2004), 243-270.








    

   
    \bibitem{Saponari_2017} Saponari, M., Boscia, D., Altamura, G., Loconsole, G., Zicca, S., D'Attoma, G., Morelli, M., Palmisano, F., Saponari, A.,
     Tavano, D., Savino, V. N., Dongiovanni, C.,  Martelli, G. P.,  Isolation and pathogenicity of {\it  Xylella fastidiosa} associated
     to the olive quick decline syndrome in southern Italy. Scientific reports,   7 (2017), 17723.
     DOI:10.1038/s41598-017-17957-z.

    \bibitem{Saponari_2018} Saponari, M., Giampetruzzi, A., Loconsole, G., Boscia, D.,  Saldarelli, P.,  {\it  Xylella fastidiosa} in olive in Apulia: Where we stand.
    Phytopathology, 109(2) (2018), 175-186.

\bibitem{Schneider_2020}	Schneider, K., van der Werf, W., Cendoya, M., Maurits, M., Navas-Cortes,
J.A., Impact of Xylella fastidiosa subspecies pauca in European olives.
PNAS, 117 (2020), 9250-9259.

    \bibitem{silva2015}  Silva  S. E., et al.,   Differential survival and reproduction in colour forms
    of {\it Philaenus spumarius} give new insights to the study
    of its balanced polymorphism. Ecological Entomology, 40 (2015), 759-766.


    
    


    \bibitem{Villalobos_2006} Villalobos, F. J., Testi, L., Hidalgo, J., Pastor, M.,  Orgaz, F. (2006). Modelling potential growth and yield of olive ({\it Olea europaea L.}) canopies.
    European Journal of Agronomy, 24(4), 296-303



    

\bibitem{white2017}  White, S. M., Bullock, J. M., Hooftman, D. A., Chapman, D. S., Modelling
 the spread and control of Xylella fastidiosa in the early stages of invasion
 in Apulia, Italy. Biological Invasions, 19(6)(2017), 1825-1837.

   
        \bibitem{Yurtsever_2000} Yurtsever, S.,  On the polymorphic meadow spittlebug,
        {\it Philaenus spumarius (L.) (Homoptera: Cercopidae)}. Turkish Journal of Zoology, 24(4) (2000), 447-460.


\end{thebibliography}
\end{document}